\documentclass[preprint2]{aastex6}
\usepackage{graphicx}
\usepackage{epstopdf}
\usepackage[section]{placeins}
\usepackage{lipsum}
\usepackage{color}
\usepackage{booktabs}
\usepackage{multirow}
\usepackage{hyperref}
\hypersetup{colorlinks=true}

\begin{document}

\title{THE EVOLUTION OF BULGE-DOMINATED FIELD GALAXIES FROM $\mathrm{z\approx1}$ TO THE PRESENT}

\author{Charity Woodrum\altaffilmark{1}}
\author{Inger J\o rgensen\altaffilmark{2}}
\author{R. Scott Fisher\altaffilmark{1}}
\author{Lindsey Oberhelman\altaffilmark{1}}
\author{Ricardo Demarco\altaffilmark{3}}
\author{Taylor Contreras\altaffilmark{1}}
\author{Jacob Bieker\altaffilmark{1}}
\altaffiltext{1}{University of Oregon, Department of Physics, 1274 University of Oregon, Eugene, OR 97403, USA}
\altaffiltext{2}{Gemini Observatory, 670 North A`ohoku Place, Hilo, HI 96720, USA}
\altaffiltext{3}{University of Concepci\'on, Department of Astronomy, Casilla 160-C, Concepci\'on, Chile}

\accepted{for publication in The Astrophysical Journal 08/20/2017}
\begin{abstract}
We analyze the stellar populations and evolutionary history of bulge-dominated field galaxies at redshifts $\mathrm{0.3<z<1.2}$ as part of the Gemini/\textit{HST} Galaxy Cluster Project (GCP). High signal-to-noise optical spectroscopy from the Gemini Observatory and imaging from the \textit{Hubble Space Telescope} are used to analyze a total of 43 galaxies, focusing on the 30 passive galaxies in the sample. Using the size-mass and velocity dispersion-mass relations for the passive field galaxies we find no significant evolution of sizes or velocity dispersions at a given dynamical mass between $\mathrm{z \approx 1}$ and the present. We establish the Fundamental Plane and study mass-to-light (\textit{M/L}) ratios. The \textit{M/L} vs. dynamical mass relation shows that the passive field galaxies follow a relation with a steeper slope than the local comparison sample, consistent with cluster galaxies in the GCP at z=0.86. This steeper slope indicates that the formation redshift is mass dependent, in agreement with ``downsizing," meaning that the low mass galaxies formed their stars more recently while the high mass galaxies formed their stars at higher redshift. The zero point differences of the scaling relations for the \textit{M/L} ratios imply a formation redshift of $\mathrm{z_{form}=1.35_{-0.07}^{+0.10}}$ for the passive field galaxies. This is consistent with the $\mathrm{(H\delta_A + H\gamma_A)'}$ line index which implies a formation redshift of $\mathrm{z_{form}=1.40^{+0.60}_{-0.18}}$.

\end{abstract}

\section{Introduction}




One of the challenging problems in modern astrophysics is to achieve a full understanding of how galaxies in the local universe acquired their observed properties. In this context, the evolution of baryons, in particular stellar mass growth, has resulted in $\sim74$\% of stars in the local universe being locked in spheroids \citep{Fukugita1998}, i.e. early-type galaxies or the bulges of late-type galaxies.


As a function of morphology, estimates of the fraction of field galaxies at $0.4 < z < 0.8$ show that late-type, disk galaxies make up $\mathrm{\sim31}$\% of the population with only $\mathrm{\sim17}$\% of galaxies being early-type. The local population of galaxies at $\mathrm{z\approx0}$ shows a sharp increase in spirals up to $\mathrm{\sim72}$\%, with most of this evolution attributed to peculiar galaxies transforming into spirals by $\mathrm{z\approx0}$. In the same redshift interval, the fraction of early-type galaxies remains almost constant. \citep{DelgadoSerrano2010}. A much milder evolution for spirals and early-type galaxies has been reported by \citet{Conselice2004}, but the difference in the result is likely due to the shallower sample considered. In high density environments, the morphological mix of galaxies differs with respect to that in the field (e.g., \citealt{Dressler1980, Holden2007, vanderWel2007, Postman2015}). The early-type galaxy fraction reaches $\mathrm{\sim80}$\% in the core of massive clusters, a fraction that seems non-evolving over the redshift range $\mathrm{0.03< z< 0.8}$ for stellar masses of the galaxies larger than $4 \cdot 10^{10} M_{\odot}$ \citep{vanderWel2007}. 


On one hand, the environment (field vs. cluster) seems to play a role in establishing galaxy properties such as stellar mass, morphology and size, although the details are still not well understood. On the other hand, recent evidence has been found that cluster and field ellipticals at $\mathrm{z\approx1.3}$ do not differ much from each other at a fixed stellar mass, although there seems to be a clear lack of massive ($\mathrm{M_* > 2\times10^{11} M_{\odot}}$) and large ($\mathrm{r_e > 4 - 5}$ kpc) field ellipticals as compared to those in clusters \citep{Saracco2017}.

\cite{Saglia2010} find that both field and cluster early-type galaxies are smaller and have higher velocity dispersions at a given dynamical mass in the redshift range $\mathrm{0.2<z<0.9}$ than at z=0. In contrast to this, no significant evolution in size or velocity dispersion at fixed dynamical mass was found for cluster galaxies from z=0.86 and z=1.27 to the present by \cite{Jorgensen2013} and \cite{Jorgensen2014}, respectively. A reason for the possible disagreement between these studies may be due to \cite{Jorgensen2013} studying higher density clusters with velocity dispersion $\mathrm{\sigma_{cluster}=1110-1450\ km\ s^{-1}}$ and \cite{Saglia2010} studying mostly lower density clusters with velocity dispersion $\mathrm{\sigma_{cluster}=300-700\ km\ s^{-1}}$. \cite{Oldham2017} find very little size and velocity dispersion evolution at a given stellar mass. However, several other studies using stellar masses find that at a given mass, galaxies are smaller at higher redshift. \citep{Trujillo2007, vanDokkum2010, Newman2012, Belli2014}


Studies of the Fundamental Plane (FP; \citealt{Dressler1987, Djorgovski1987, Jorgensen1996}) of cluster galaxies with masses $>10^{11} M_{\odot}$ at intermediate redshifts have shown that the evolution of the galaxy \textit{M/L} is consistent with passive evolution (e.g., \citealt{Jorgensen2006, Jorgensen2007}; \citealt{vanDokkum2007}). The offsets in the \textit{M/L} ratio with respect to Coma imply formation redshifts for massive ($\mathrm{M\gtrsim 10^{11}M_{\odot}}$) cluster early-type galaxies of $\mathrm{z_{form}\approx 2}$, which corresponds to somewhat older ($\mathrm{\sim0.4\ Gyr}$) stellar populations in those galaxies compared with field counterparts from the same study \citep{vanDokkum2007}.  This difference is almost negligible and is consistent with the similar cluster and field formation epochs reported by \cite{Rettura2011}. Moreover, it has been found that the FP is not significantly different for field and cluster galaxies at z=0 (e.g., \citealt{Treu2001} and references therein).

A steeper FP at higher redshift and relatively younger ages for massive cluster early-type galaxies at intermediate-to-high redshifts have been found by \cite{Jorgensen2006, Jorgensen2007} and \cite{Jorgensen2013}. Similarly, a steeper FP at higher redshift for $\mathrm{z\sim1}$ field early-type galaxies has also been reported by \cite{diSeregoAlighieri2006err, diSeregoAlighieri2006}, \cite{Treu2005}, and \cite{vanderWel2005}. A steeper slope of the FP is understood to be the result of ``downsizing," where high mass galaxies formed their stars at higher redshifts while low mass galaxies formed their stars more recently.


Line indices may be used to determine ages, metallicities, and element abundance ratios of stellar populations in galaxies and ultimately the evolution of early-type galaxies in clusters (e.g., \citealt{Jorgensen2005, Jorgensen2013}) and in the field (e.g., \citealt{Ziegler2005, Schiavon2006, Choi2014, Gallazzi2014}). 
 Studies of the line indices in quiescent field galaxies show no significant evolution in metallicities and are consistent with passive evolution since $\mathrm{z\sim0.7}$ \citep{Choi2014, Gallazzi2014}. Studies of line indices in early-type galaxies such as \cite{Harrison2011} and \cite{SanchezBlazquez2006iii, SanchezBlazquez2006i, SanchezBlazquez2006ii} have found that absorption-line strengths, stellar population parameters, age, and metallicity do not differ between galaxies in the cluster centers and their outskirts. In contrast to this, \cite{Thomas2005} found a $\sim2$ Gyr difference in age between massive early-type galaxies in the field and massive early-type galaxies in clusters in the local universe. Also, \cite{Choi2014} find tentative evidence for cluster galaxies containing older stellar populations than field galaxies at similar redshifts. However, \cite{Thomas2010} find no significant difference between galaxies in the field and in clusters.
 
 Motivated by the lack of consensus on the possible age difference between passive field and cluster galaxies and by a paucity of published work that treats this problem simultaneously with both high-quality spectra and accurate photometry, we address this issue using the rich spectrophotometric dataset of the Gemini/\textit{HST} Galaxy Cluster Project (GCP; \citealt{Jorgensen2005, Jorgensen2013}). Although the GCP sample is small compared to forthcoming results from the LEGA-C survey by \cite{vanderWel2016}, its high resolution imaging and high signal-to-noise (S/N) spectroscopic data allow for precise measurements of velocity dispersions, line indices, dynamical masses, luminosities and sizes. This allows us to derive and compare the formation redshifts for field and cluster galaxies and analyze the properties of the FP for those samples.
 
 This paper is organized as follows: \S \ref{data} describes our observational data and sample selection process, \S \ref{methods} introduces the methods, models and scaling relations used, and \S \ref{results}, \S \ref{discussion}, and \S \ref{conclusions} describe the results, discussion and conclusions, respectively.

Throughout this paper we adopt a $\mathrm{\Lambda}$CDM cosmology with $\mathrm{H_0=70 \ km\ s^{-1} \ Mpc^{-1}}$, $\mathrm{\Omega_M = 0.3}$,  $\mathrm{\Omega_{\Lambda} =0.7}$, and all apparent magnitudes are in the AB system \citep{OkeGunn1983}. 
\section{Observational Data}\label{data}
\subsection{Gemini North}

The field galaxies we analyze here were observed serendipitously with the galaxy clusters MS0451.6-0305Â (z=0.54), RXJ0152.7-1357 (z=0.83), and RXJ1226.9+3332 (z=0.89), which are part of the GCP. The data for cluster galaxies in MS0451.6-0305, RXJ0152.7-1357, RXJ1226.9+3332, are analyzed and published in \cite{Jorgensen2005} and \cite{Jorgensen2013}. 

All spectra of the field galaxies in this paper were obtained with Gemini North using the Gemini Multi-Object Spectrograph (GMOS-N) \citep{Hook2004} in the multi-object spectroscopic mode.  The spectroscopic sample selection for this paper is based on GMOS-N imaging using \textit{r'}, \textit{i'}, and \textit{z'} filters. Galaxies were classified as expected cluster members based on their colors. For galaxies imaged in the field of cluster MS0451.6-305, (\textit{r'-z'}) and (\textit{i'-z'}) colors were used. For galaxies in the field of cluster RXJ0152.7-1357, (\textit{i'-z'}) was used. For galaxies in the field of cluster RXJ1226.9+3332, (\textit{r'-z'}), (\textit{i'-z'}), and (\textit{r'-i'}) colors were used.  When mask space was not filled with assumed cluster members, expected non-members were added. Our sample of field galaxies was selected from the observed galaxies that were determined to be non-members. For a detailed description on the methods of the spectroscopic sample selection refer to \cite{Jorgensen2005} and \cite{Jorgensen2013}.

This is the first analysis of the data for the 82 field galaxies and the first publication of their spectra. The data tables, greyscale images and sample spectra of the field galaxies can be found in Appendix \ref{appendixA}. Since the field and cluster galaxy data were obtained simultaneously, they were reduced in the same manner, the details of which are published in \cite{Jorgensen2005} and \cite{Jorgensen2013}. The median S/N per $\mathrm{\AA}$ in the rest-frame of the 82 field galaxies is S/N=28, while the 30 passive bulge-dominated field galaxies have a median S/N=35.

From the spectra of the galaxies, we measure redshifts, velocity dispersions and line indices. In this paper we use a subset of the Lick/IDS absorption line indices \citep{Worthey1994}, see the data tables and sample spectra in Appendix \ref{appendixA}. Considering the redshift intervals spanned by our spectra, indices blueward of $\mathrm{4700\ \mathring{A}}$ in the rest-frame are available across our whole sample of field galaxies ($\mathrm{0.3<z<1.2}$) and for all of our comparison sources from Perseus and Abell 194 ($\mathrm{z=0.02}$) to RXJ1226.9+3332 ($\mathrm{z=0.89}$).

Specifically, we use the higher order Balmer line indices $\mathrm{H\delta_A}$ and $\mathrm{H\gamma_A}$ \citep{Worthey1997} as well as the H$\zeta_A$ index \citep{Nantais2013} as primary indicators of the ages of the stellar populations. In our analysis, we use the combined index $\mathrm{(H\delta_A + H\gamma_A)'}$ as defined in \cite{Kuntschner2000}, namely:
\begin{equation}
\mathrm{(H\delta_A + H\gamma_A)' \equiv -2.5log(1-\frac{H\delta_A + H\gamma_A}{43.75+38.75})}\ .
\end{equation}

As metallicity indicators we use the indices CN3883 \citep{Davidge1994}, Fe4383 \citep{Worthey1994}, and C4668 \citep{Worthey1994, Tripicco1995}. 

We adopt the typical uncertainties on the line indices from Table 19 in \cite{Jorgensen2013}, which were estimated from internal comparisons of the line indices. This is possible because the field sample was observed at the same time as the cluster sample, guaranteeing similar observing conditions and the same observational setup. For a detailed explanation on the calculation of the uncertainties, we refer the reader to \cite{Jorgensen2013}.

\subsection{Hubble Space Telescope } 
The cluster and field galaxies in this paper were observed with the \textit{Hubble Space Telescope's} (\textit{HST}) Advanced Camera for Surveys (ACS) and the data were obtained from the \textit{HST} archive. The photometric parameters for our sample of field galaxies are published in \cite{Chiboucas2009} and \cite{Jorgensen2013}. They used the fitting program GALFIT \citep{Peng2002} to derive the mean surface brightnesses, effective radii, and total magnitudes. The magnitudes were calibrated to rest-frame B using colors from the ground based imaging and the same technique as described in \cite{Jorgensen2013}, see also \cite{Blanton2003}. The S\'ersic profile \citep{Sersic1968} was used to determine the S\'ersic index, $\mathrm{n_{ser}}$, to select bulge-dominated galaxies. In our investigation of the FP, we use parameters that were fit using a $\mathrm{r^{1/4}}$ profile \citep{deVaucouleurs1948} to be consistent with the analysis of our low redshift comparison sample. In Section \ref{fpsection} we present a brief test on the possible bias of this choice.

\begin{figure}[h!]
	\epsscale{1.1}
	\plotone{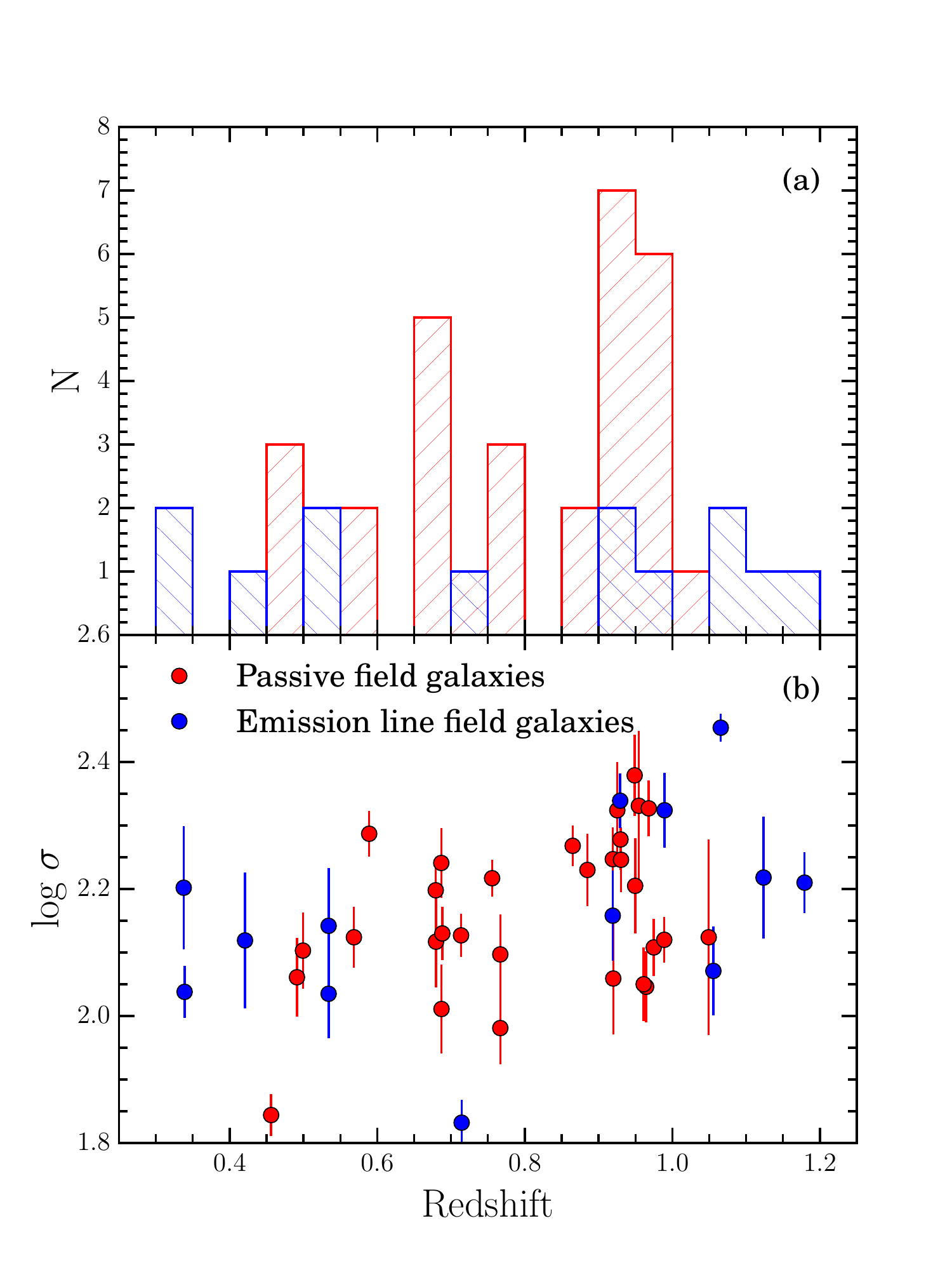}
	\caption{(a) Redshift distributions of the two samples with a 0.05 bin size, passive field galaxies in red and emission line field galaxies in blue. (b) Velocity dispersion versus redshift. This figure is shown to display the redshift and velocity dispersion distributions for the passive and emission line field galaxies in our samples. The apparent clustering around $\mathrm{z\approx 0.95}$ is due to selection effects, not actual clustering. \label{zdistribution}}\end{figure}

\subsection{Field Galaxy Sample} \label{sample}

From a total of 82 field galaxies with spectroscopy, we select our passive field galaxy sample with the same criteria as those in \cite{Jorgensen2013} to be consistent with the cluster galaxy comparison samples. The selection criteria are the following (also see Table \ref{samples}):
\begin{enumerate}
	\item Spectroscopy with $\mathrm{S/N \geq 20\ per\ \mathring{A}}$ in the rest frame
	\item A redshift cutoff of $\mathrm{z \geq 0.3}$ to exclude a few low redshift galaxies from our intermediate redshift sample
	\item A S\'ersic index of $\mathrm{n_{ser} \geq 1.5}$ \citep{Sersic1968} to select for bulge-dominated galaxies
	\item $\mathrm{EW[O\ \textsc{ii}]\leq 5 \mathring{A}}$ to exclude emission line galaxies, which may be star forming or contain an active galactic nuclei
\end{enumerate}

Throughout this paper, the plots contain both the passive field galaxies and emission line field galaxies, though we concentrate on the passive field galaxy sample to allow a more consistent comparison between the field galaxies and the cluster galaxies in \cite{Jorgensen2013}. To provide approximately equally sized subsamples of passive field galaxies, we split our sample into two redshift bins, $\mathrm{0.3<z<0.8}$ and $\mathrm{0.8<z<1.2}$, with median redshifts of $\mathrm{z=0.68}$ and $\mathrm{z=0.95}$. Henceforth, we will refer to the galaxies in the lower redshift range as the $\mathrm{z\approx0.7}$ sample, and galaxies in the higher redshift range as the $\mathrm{z\approx1}$ sample. The redshift distributions and velocity dispersion distributions for the passive and emission line field galaxies are shown in Figure \ref{zdistribution}. Throughout the paper, the number of field galaxies varies between plots because not all of the field galaxies have measurements for all line indices.

\subsection{Local Comparison Sample}
The Coma, Perseus, and Abell 194 clusters make up our local comparison sample. Perseus and Abell 194 are used for the analysis of line indices, while Coma is used for the size-mass, size-velocity dispersion, \textit{M/L}-mass, and \textit{M/L}-velocity dispersion relations as well as the derivation of the FP. Spectroscopic data for the Coma cluster (z=0.024) are published and detailed in \cite{Jorgensen1999}, while the spectroscopic data for Perseus (z=0.018) and Abell 194 (z=0.018) are described in \cite{Jorgensen2005} and will be published in J\o rgensen et al. (in preparation). The same B-band photometry was used for the Coma cluster as in \cite{Jorgensen2005}. All three samples were filtered to exclude spiral, irregular, emission line, and low mass ($\mathrm{< 10^{10.3} M_{\odot}}$) galaxies in order to best compare with our field galaxy sample.
\begin{deluxetable}{ccccc}
	
	
	
	
	\tablecaption{Field Galaxy Selection Criteria\label{samples}}
	
	
	\tablehead{\colhead{} & \colhead{$\mathrm{N_{gal}}$} & \colhead{$\mathrm{n_{ser}}$} & \colhead{EW[O \textsc{ii}]} & \colhead{} 
	} 
	
	\startdata
	\multicolumn{1}{l}{Total} & 82 &  &  \\ 
	\multicolumn{1}{l}{Disk-dominated} & 30 & $\mathrm{< 1.5}$ &  \\ 
	\multicolumn{1}{l}{Bulge-dominated} &  &  &  \\
	\multicolumn{1}{l}{\ \ \ \ Emission line\tablenotemark{a}} & 13 & $\mathrm{\geq 1.5}$ & $\mathrm{> 5\AA}$ \\
	\multicolumn{1}{l}{\ \ \ \ Passive} & 30 & $\mathrm{\geq 1.5}$ & $\mathrm{\leq 5\AA}$ \\ 
	\multicolumn{1}{l}{Excluded\tablenotemark{b}} & 9 &  &  &  \\
	\enddata
	
	\tablenotetext{a}{When the spectra do not cover the [O \textsc{ii}] emission lines, we use $\mathrm{H\beta}$ and $\mathrm{[O\ \textsc{iii}]}$ as an indication of star formation.}
	\tablenotetext{b}{Field galaxies that did not meet one or more criteria described in the text.}

	
\end{deluxetable}

\section{Methods} \label{methods}
The methods used to analyze the field galaxy sample are established in \cite{Jorgensen2013} and briefly summarized here.

\subsection{Stellar Population Models} \label{models}
The interpretation of the line index values was carried out using the same models as done by \cite{Jorgensen2013}. This allows for a consistent comparison between their cluster galaxy data and our field galaxy data. 

Our analysis is based on the single stellar population (SSP) models from \cite{Thomas2011} and \cite{MarastonandStromback2011}. We use the models that assume a \cite{Salpeter1955} initial mass function. The output of the \cite{Thomas2011} models are line indices for ages of 1-15 Gyr, a range in metallicity with specific values of $\mathrm{[M/H]=-0.33, 0.00, 0.35, 0.67}$, and a range in $\alpha$-element abundance with ratios $\mathrm{[\alpha/Fe]=0.0, 0.3, 0.5}$. The output of the \cite{MarastonandStromback2011} models are spectra with ages of 1-15 Gyr, a range in metallicity with [M/H]=-0.3, 0.0, 0.3 and solar abundance ratio $\mathrm{[\alpha/Fe]=0}$. As it is customary, sets of age, [M/H], $\mathrm{[\alpha/Fe]}$ were used to create model grids for pairs of line indices that were then directly compared with the measured index values. 

The CN3883 index has not been modeled by any of the commonly available models in the literature which also take into account non-solar $\mathrm{[\alpha/Fe]}$. Therefore, we follow the approach by \cite{Jorgensen2013} and adopt their Equation 4 (see also their Fig. 5 and Table 9) that relates $\mathrm{CN_2}$ with CN3883. Then we used that equation and the $\mathrm{CN_2}$ values from \cite{Thomas2011} to obtain CN3883 model estimates. $\mathrm{H\zeta_A}$ were derived from the \cite{MarastonandStromback2011} model spectra and are only available for [$\mathrm{\alpha/Fe}$]=0. Finally, we use \textit{M/L} ratios modeled by \cite{Maraston2005} with solar abundance ratios as the \cite{Thomas2011} models do not provide \textit{M/L} predictions.
\begin{deluxetable*}{lrrrrrrrrr}
	\tabletypesize{\footnotesize}
	\tablecaption{Scaling Relations\label{scalingrelations}}
	\tablehead{\multicolumn{1}{c}{\multirow{2}[3]{*}{Relation}} &
		\multicolumn{3}{c}{Local Sample} &
		\multicolumn{3}{c}{$\mathrm{z\approx0.7}$ Passive Field} &
		\multicolumn{3}{c}{$\mathrm{z\approx1}$ Passive Field} 
		\\
		\cmidrule(lr){2-4}\cmidrule(lr){5-7} \cmidrule(lr){8-10} &
		\colhead{$\mathrm{\gamma}$} & \colhead{$\mathrm{N_{gal}}$} & \colhead{rms} & \colhead{$\mathrm{\gamma}$} & \colhead{$\mathrm{N_{gal}}$} & \colhead{rms} & \colhead{$\mathrm{\gamma}$} & \colhead{$\mathrm{N_{gal}}$} & \colhead{rms} \\ 
		\colhead{(1)} & \colhead{(2)} & \colhead{(3)} & \colhead{(4)} & \colhead{(5)} & \colhead{(6)} & \colhead{(7)} & \colhead{(8)} & \colhead{(9)} & \colhead{(10)} } 
	\startdata
	$\mathrm{log\ r_e = (0.57 \pm\ 0.06)\ log\ Mass + \gamma}$ & $-5.73^{+0.16}_{-0.15}$ & 105 & 0.16 & $-5.75^{+0.17}_{-0.07}$  & 14 & 0.20 & $-5.77^{+0.17}_{-0.15}$  & 16 & 0.14 \\
	$\mathrm{log\ \sigma = (0.26 \pm 0.03)\ log\ Mass + \gamma}$ & $-0.67^{+0.07}_{-0.07} $ & 105 & 0.08 & $-0.66^{+0.04}_{-0.07} $ & 14 & 0.10 & $-0.64^{+0.06}_{-0.10} $ & 16 & 0.07 \\
	$\mathrm{log\ \textit{M/L} = (0.24 \pm 0.03)\ log\ Mass + \gamma}$ & $-1.75^{+0.08}_{-0.10}$  & 105 & 0.09 & $-2.29^{+0.16}_{-0.40}$  & 14 & 0.24 & $-2.47^{+0.33}_{-0.22}$  & 16 & 0.25 \\
	$\mathrm{log\ \textit{M/L} = (1.07 \pm 0.12)\ log\ \sigma + \gamma}$ & $-1.56^{+0.14}_{-0.08}$  & 105 & 0.11 & $-2.10^{+0.26}_{-0.35}$  & 14 & 0.25 & $-2.23^{+0.27}_{-0.23}$  & 16 & 0.23 \\
	$\mathrm{(H\delta_A + H\gamma_A)' = (-0.085 \pm 0.015)\ log\ \sigma + \gamma}$ & $0.10^{+0.01}_{-0.01} $  & 65 & 0.02 & $0.16^{+0.05}_{-0.02}$  & 9 & 0.04 & $0.19^{+0.06}_{-0.06} $ & 8 & 0.08 \\
	$\mathrm{CN3883 = (0.29 \pm 0.06)\ log\ \sigma + \gamma}$ & $-0.41^{+0.05}_{-0.05}$  & 65 & 0.05 & $-0.43^{+0.07}_{-0.05}$  & 10 & 0.05 & $-0.45^{+0.06}_{-0.03}$  & 14 & 0.05 \\
	$\mathrm{log\ Fe4383 = (0.19 \pm 0.07)\ log\ \sigma + \gamma}$ & $0.26^{+0.05}_{-0.07}$  & 65 & 0.06 & $0.11^{+0.21}_{-0.05}$ & 12 & 0.18 & $0.07^{+0.27}_{-0.22}$ & 10 & 0.27 \\
	$\mathrm{log\ C4668 = (0.33\ \pm 0.08)\ log\ \sigma + \gamma}$ & $0.11^{+0.06}_{-0.04}$ & 65 & 0.06 & $0.03^{+0.10}_{-0.20}$ & 10 & 0.16 & $0.01^{+0.05}_{-0.00}$ & 3 & 0.03 \\
	$\mathrm{log\ H\zeta_A = (-0.76\ \pm 0.29)\ log\ \sigma + \gamma}$ & $1.76^{+0.21}_{-0.21}$ & 45 & 0.25 & $2.00^{+0.20}_{-0.17}$ & 4 & 0.14 & $1.96^{+0.27}_{-0.08}$ & 13 & 0.18 \\
	$\mathrm{log\ r_e = (1.3 \pm 0.08)\ log\ \sigma - (0.82 \pm 0.03)\ log <I>_e + \gamma}$ & $-0.44^{+0.06}_{-0.08}$ & 105 & 0.08 & $0.00^{+0.12}_{-0.35}$ & 14 & 0.20 & $0.15^{+0.27}_{-0.19}$ & 16 & 0.21 \\
	\\
	\textbf{RXJ0152.7-1357, RXJ1226.9+3332\tablenotemark{a}} &  &  &  &  &  &  &  &  &  \\
	$\mathrm{log\ \textit{M/L} = (0.55 \pm 0.08)\ log\ Mass + \gamma}$ & $-5.85^{+0.25}_{-0.10}$ & 49 & 0.14 & $-5.70^{+0.25}_{-0.15}$ & 14 & 0.21 & $-5.90^{+0.32}_{-0.12}$ & 16 & 0.20 \\
	\enddata
	\tablenotetext{a}{The slope derived from a fit treating RXJ0152.7-1357 and RXJ1226.9+3332 as one sample}
	\tablecomments{Column 1: scaling relation for the local comparison samples unless otherwise noted; Column 2: zero point for the low-redshift sample,\\ uncertainties are calculated with bootstrap resampling; Column 3: number of galaxies included from the low-redshift sample; Column 4:\\ rms in the Y-direction of the scaling relation for the low-redshift sample; Columns 5 - 7: zero point, number of galaxies, and rms in the \\Y-direction for the $\mathrm{z\sim0.7}$ passive field galaxies; Columns 8 - 10: zero point, number of galaxies, and rms in the Y-direction for the $\mathrm{z\approx1}$ \\passive field galaxies.}
	\tablerefs{Columns 1-4 adopted from \cite{Jorgensen2013}, except the relation for $\mathrm{log\ H\zeta_A}$ which is from\\ \cite{Jorgensen2014}}.
	
\end{deluxetable*}

\subsection{Adopted Scaling Relations}
We compare the field galaxies to cluster galaxies at similar redshifts and to cluster galaxies at $\mathrm{z\approx 0}$ by fitting scaling relations to the samples. We assume that the relations are linear in log-space, since previous studies of larger samples in clusters do not find any non-linearities, except in the log $\mathrm{\sigma}$ vs. log Mass relation for galaxies with masses above $\mathrm{10^{12}M_{\odot}}$ \citep{Jorgensen2013, Jorgensen2014}. We use the same fitting technique as \cite{Jorgensen1996}, which minimizes the sum of the absolute residuals perpendicular to the relation. This technique is robust to outliers and does not assume any particular distribution of the residuals. The uncertainties of the coefficients are determined with a bootstrap method. Fits to the field galaxy samples show that the slopes of the relations are consistent at the $\mathrm{1\sigma}$ level with the better determined cluster galaxy relations \citep{Jorgensen2013, Jorgensen2014}. We therefore adopt the scaling relations from \cite{Jorgensen2013}, except for the scaling relation for $\mathrm{H\zeta_A}$ which is derived in \cite{Jorgensen2014}. These relations are listed in Table \ref{scalingrelations}, Column 1. The median zero points of the three samples are listed in Table \ref{scalingrelations}, Columns 2, 5 and 8. The uncertainties on these values are calculated using bootstrap resampling.

We chose RXJ1226.9+3332 as the high redshift comparison cluster because the X-ray structure of this cluster is the most relaxed of the high redshift clusters in \cite{Jorgensen2013}. The smooth, symmetric X-ray emission implies that RXJ1226.9+3332 has not undergone a recent cluster-cluster merger interaction, as opposed to the RXJ0152.7-1357 cluster at $\mathrm{z=0.84}$, which shows an elongated X-ray morphology indicative of an ongoing merger \citep{Girardi2005, Jorgensen2005, Demarco2005, Demarco2010}.

We then derive the median zero point differences, $\mathrm{\Delta \gamma}$, between the field galaxy samples and the local comparison sample. The random uncertainties on these, $\mathrm{\sigma_{\Delta \gamma}}$, are derived using two methods: (1) paired bootstrap resampling and (2) using the following equation:
\begin{equation}
\label{randomuncertainty}
\mathrm{\sigma_{\Delta\gamma}=(rms_{local}^2/N_{local} + rms_{FG}^2/N_{FG})^{0.5}},
\end{equation}
where subscripts ``FG" refer to the field galaxy sample and ``local" refer to the $\mathrm{z\approx 0}$ cluster sample. These methods produce similar results and we therefore use the values derived from Equation \ref{randomuncertainty} throughout the text and figures.

In the following, we compare the scaling relation zero points for the two passive field galaxy samples to the local comparison sample. For the \textit{M/L}-mass relation, we also compare the passive field galaxies to the cluster galaxies at high redshift. \cite{Jorgensen2013} fit RXJ0152.7-1357 and RXJ1226.9+3332 separately and determined that the slopes were consistent with one another. We therefore use the relation of these two clusters treated as one sample to compare to the passive field galaxies. The zero points and scatter are summarized in Table \ref{scalingrelations}.

\section{Results} \label{results}
\
\subsection{Effective Radius and Velocity Dispersion  vs. Dynamical Mass}

\begin{figure}[h!]
	\epsscale{1.1}
	\plotone{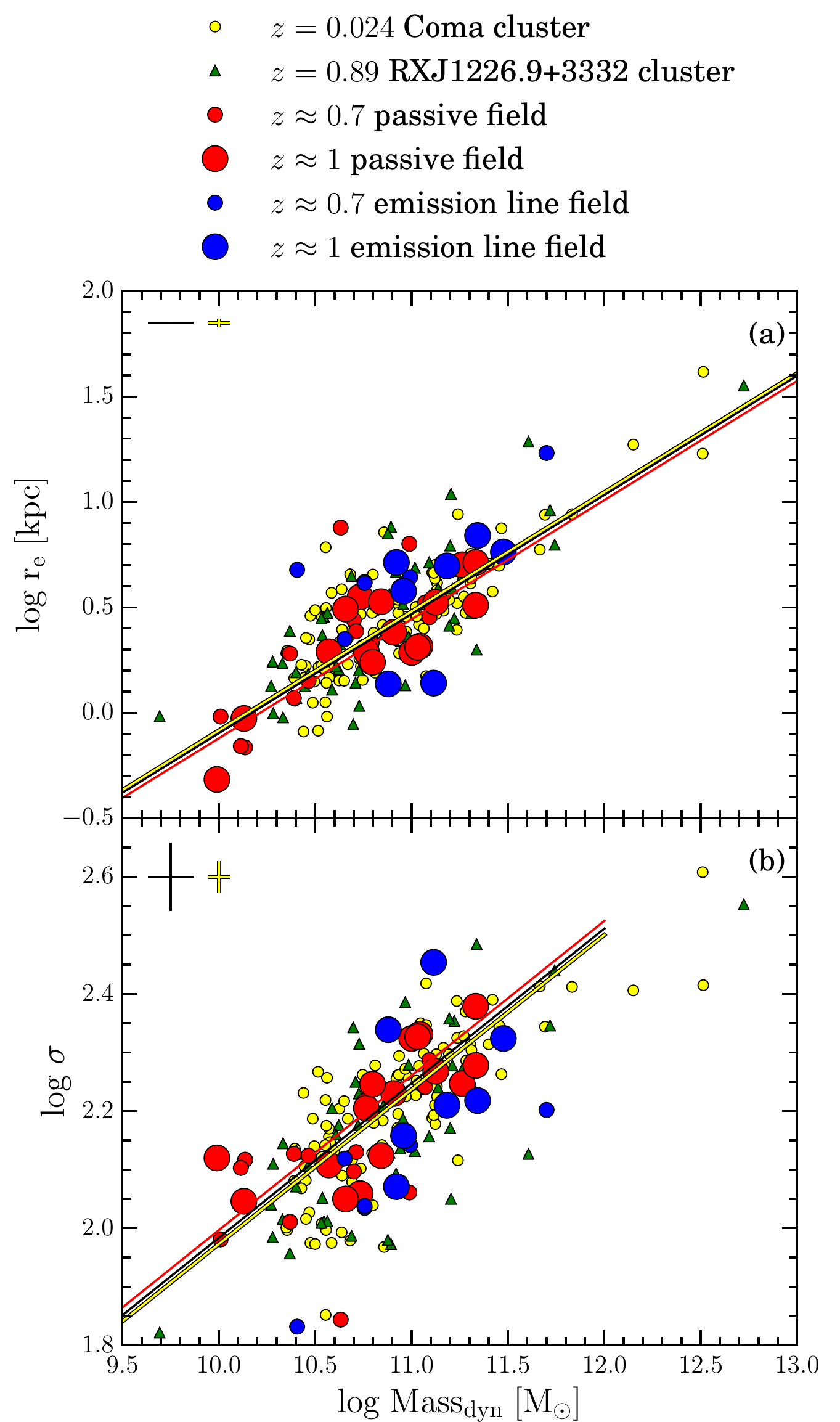}
	\caption{(a) Effective radius versus dynamical galaxy mass. (b) Velocity dispersion versus dynamical galaxy mass. Yellow circles: Coma cluster galaxies. Green triangles: RXJ1226.9+3332 $\mathrm{(z=0.89)}$ cluster galaxies. Small red circles: $\mathrm{z\approx0.7}$ passive field galaxies. Large red circles: $\mathrm{z\approx1}$ passive field galaxies. Small blue circles: $\mathrm{z\approx0.7}$ emission line field galaxies. Large blue circles: $\mathrm{z\approx1}$ emission line field galaxies. Yellow line: best fit relation for the Coma cluster \citep{Jorgensen2013}.  Black line and red line: best fit relation for Coma cluster offset to the median zero points of the  $\mathrm{z\approx0.7}$ passive field galaxies and $\mathrm{z\approx1}$ passive field galaxies, respectively. Black error bars display the typical uncertainties for the passive field and RXJ1226.9+3332 galaxies, yellow error bars display the typical uncertainties for the local comparison sample. The passive field galaxies follow the same relations as found for the Coma cluster sample. We therefore conclude that at a given dynamical mass, the passive field galaxies show no significant evolution of size or velocity dispersion between $\mathrm{z \approx 1}$ and the present.\label{lrelsigmalmass}}
\end{figure}

Figure \ref{lrelsigmalmass}a shows the effective radius versus dynamical galaxy mass and Figure \ref{lrelsigmalmass}b shows the velocity dispersion versus dynamical galaxy mass. The galaxy masses are derived using the approximation $\mathrm{Mass=5r_{e}\sigma^{2}G^{-1}}$ \citep{Bender1992}. We assume that field and cluster galaxies at z $\approx0$ have identical size-mass relations \citep{HuertasCompany2013}. We then compare the slopes of the size-mass and velocity dispersion-mass relations between the local comparison sample and the passive field galaxies and find no significant difference. We therefore adopt the slope of the local relation for the field galaxy sample and find the median offsets in the zero points for the passive field galaxies, see Table \ref{scalingrelations}. The zero point differences with respect to Coma in $\mathrm{log\   r_e}$ for the $\mathrm{z\approx0.7}$ and $\mathrm{z\approx1}$ passive field galaxy samples are $\mathrm{\Delta log\ r_e= -0.02\pm0.06}$ and $-0.04\pm0.04$, respectively, and therefore not significant. The zero point differences with respect to Coma in $\mathrm{log\   \sigma}$ for the $\mathrm{z\approx0.7}$ and $\mathrm{z\approx1}$ passive field galaxies are $\mathrm{\Delta log\ \sigma= 0.01\pm 0.03\ and\ 0.02\pm0.02}$, respectively, and therefore also not significant. The random uncertainties were calculated using Equation \ref{randomuncertainty}. The passive field galaxies follow the same relations as found for the Coma cluster sample and so we conclude that at a given dynamical mass, the passive field galaxies show no significant evolution of size or velocity dispersion between $\mathrm{z \approx 1}$ and the present.

The zero point differences with respect to RXJ1226.9+3332 in log $\mathrm{r_e}$ for the $\mathrm{z\approx0.7}$ and $\mathrm{z\approx1}$ passive field galaxy samples are $\mathrm{\Delta log\ r_e= -0.02\pm0.06}$ and $\mathrm{-0.05\pm0.05}$, respectively, and therefore not significant. The zero point differences with respect to RXJ1226.9+3332 in $\mathrm{log\   \sigma}$ for the $\mathrm{z\approx0.7}$ and $\mathrm{z\approx1}$ passive field galaxies are $\mathrm{\Delta log\ \sigma= 0.02\pm0.03\ and\ 0.04\pm0.02}$, respectively, and therefore also not significant. We conclude that the passive field galaxies follow the same size-mass and velocity dispersion-mass relations as cluster galaxies at similar redshifts.

 \begin{figure}[h!]
 	\epsscale{1.3}
 	\plotone{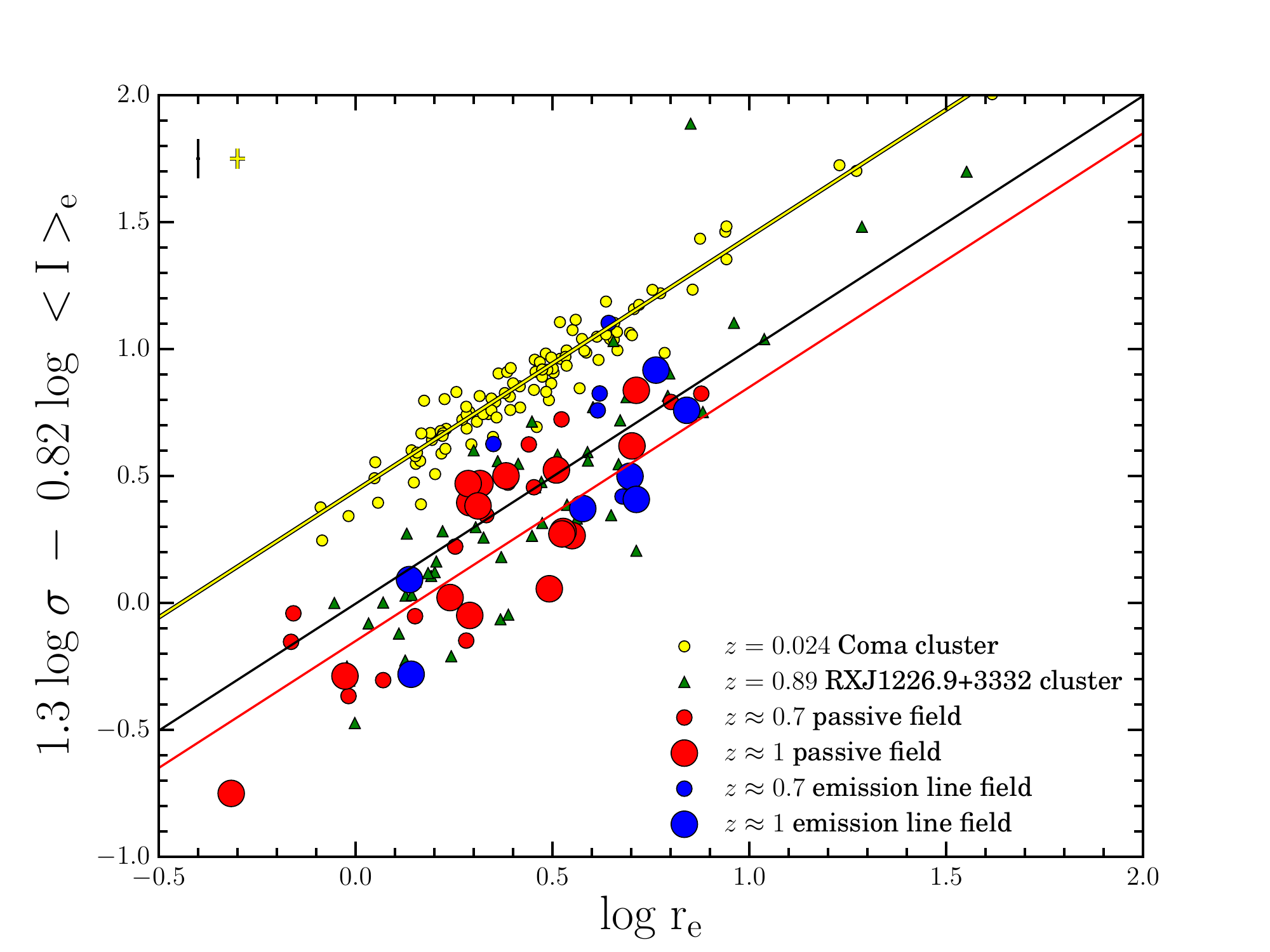}
 	\caption{The Fundamental Plane. Symbols and lines as on Figure \ref{lrelsigmalmass}. The FP for the passive field galaxies is offset and steeper than the Coma relation.\label{FP}}
 \end{figure}

\subsection{The Fundamental Plane}\label{fpsection}
\begin{figure}[h!]
  	\epsscale{1.25}
  	\plotone{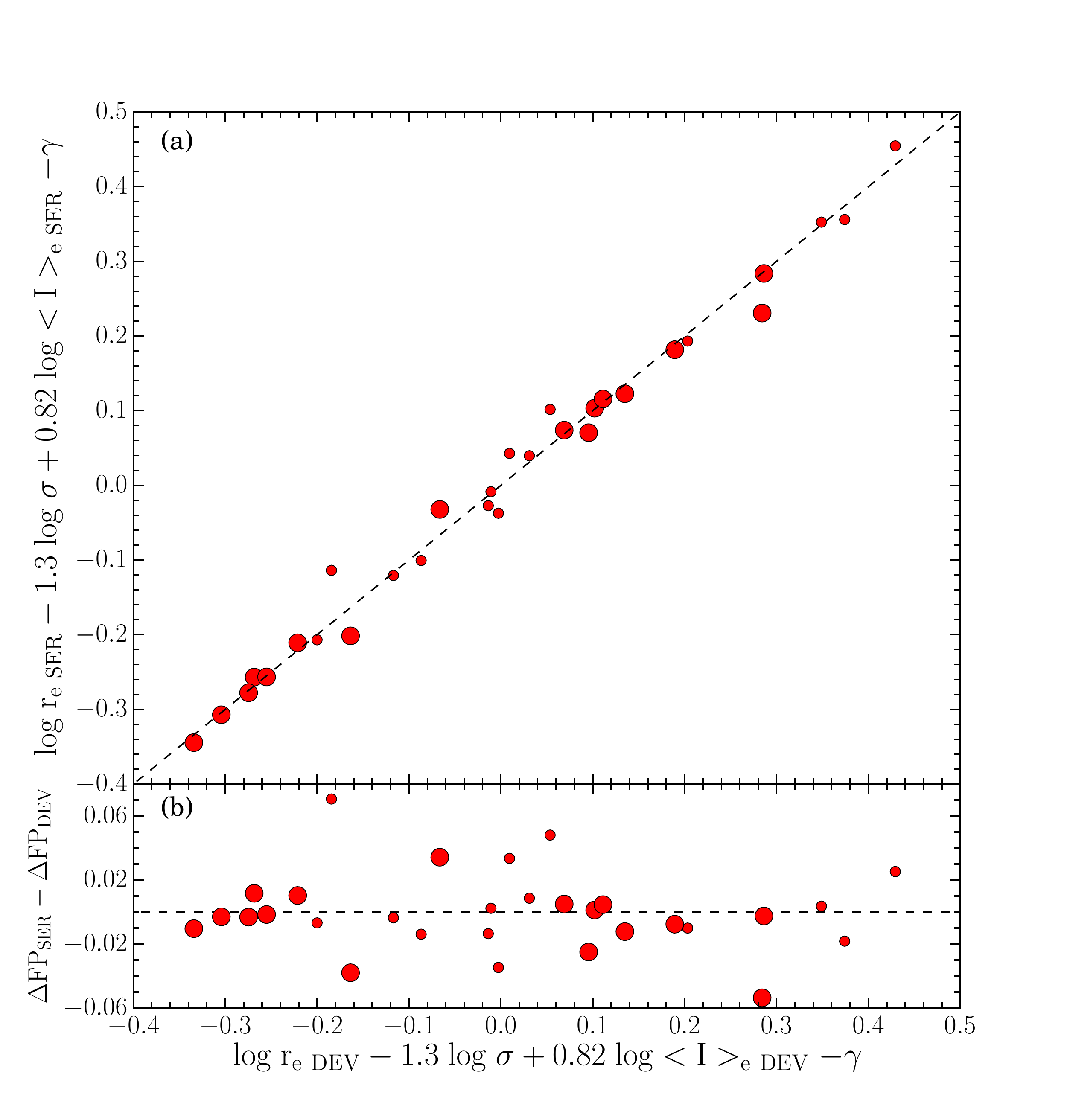}
  	\caption{(a) Residuals of the FP determined from the parameters fit with the S\'ersic profile versus those fit from the $\mathrm{r^{1/4}}$ profile. Symbols as on Figure \ref{lrelsigmalmass}. Dotted line shows a one-to-one relation. (b) The difference between the residuals of the FP determined from the two different profiles versus the residuals of the FP determined from the $\mathrm{r^{1/4}}$ profile.\label{FPserdev}}\end{figure}

Figure \ref{FP} shows the FP edge-on. The parameters for the FP are fit using the $\mathrm{r^{1/4}}$ profile. To test the validity of this choice, we investigate the residuals of the FP for the passive field galaxies determined from the parameters fit with the $\mathrm{r^{1/4}}$ profile versus those fit from the S\'ersic profile, see Figure \ref{FPserdev}. Our results do not significantly depend on our choice to use the parameters fit with the $\mathrm{r^{1/4}}$ profile. We investigate the correlation between $\mathrm{n_{ser}}$ and the residuals of the FP determined from the $\mathrm{r^{1/4}}$ profile and find no correlation. We therefore consistently use $\mathrm{r^{1/4}}$ profiles to compare between samples.

Figure \ref{lmllmasslsigma} shows a projection of the FP using \textit{M/L} ratios vs. the dynamical masses and the velocity dispersions of the galaxies. Figure \ref{lmllmasslsigma} shows the predictions based on models from \citet{Thomas2005} for the relations of the median redshifts of the two passive field galaxy samples and of the RXJ1226.9+3332 cluster. These models assume that the low-mass galaxies formed their stars more recently while the high-mass galaxies formed their stars at higher redshifts. The passive field galaxies roughly follow the models, however, they show more evolution in the \textit{M/L} vs. velocity dispersion relation than the models. The slope of the FP for the passive field galaxies is steeper than the Coma relation and consistent with the slope found for the cluster galaxies in \cite{Jorgensen2013} at $\mathrm{z\approx0.86}$.

 We use the same method as \cite{Jorgensen2013} to evaluate the evolution of the field galaxies as a function of mass and redshift. For this purpose, we divide the $\mathrm{z\approx1}$ passive field galaxies into two subsamples, one with velocity dispersions of $\mathrm{log\ \sigma < 2.24}$ and one with velocity dispersion of $\mathrm{log\ \sigma \geq 2.24}$. These two subsamples have average velocity dispersions of $\mathrm{log\ \sigma = 2.12\ and\ 2.30}$, respectively. The $\mathrm{z\approx0.7}$ field galaxy sample contains only three galaxies with $\mathrm{log\ \sigma \geq 2.24}$ (Figure \ref{zdistribution}), and therefore was not analyzed in this way.
 
 We use the equation $\mathrm{\Delta log\ \textit{M/L}=0.935 \Delta log\ age}$ \citep{Jorgensen2013} based on the SSP models in \cite{Maraston2005} to find the predicted formation redshift of the passive field galaxies under the assumption of passive evolution. We determine formation redshifts with $\mathrm{\chi^2}$ fits to the zero points differences in the \textit{M/L} vs. Mass relation for the sample of passive field galaxies, which implies $\mathrm{z_{form}=1.35^{+0.10}_{-0.07}}$. We also determine formation redshifts for the median zero point differences of the $\mathrm{z\approx 0.7}$ and $\mathrm{z\approx 1}$ samples. These results are listed in Table \ref{tab-formationredshift}. For the $\mathrm{z\approx 1}$ sample, the zero point differences of the low velocity dispersion subsample imply a $\mathrm{z_{form}=1.24^{+0.10}_{-0.07}}$ and for the high velocity dispersion subsample a $\mathrm{z_{form}=2.16^{+0.81}_{-0.39}}$.
 
 To study the intrinsic scatter relative to the FP, we adopt the slope for the log \textit{M/L} vs. log Mass relation from the RXJ1226.9+3332 and RXJ0152.7-1357 cluster galaxies treated as one sample and then calculate the median offsets in the zero points for the $\mathrm{z\approx0.7}$ and $\mathrm{z\approx1}$ passive field galaxies, see Table \ref{scalingrelations}. We then calculate the scatter with respect to this offset relation for the passive field galaxies. To calculate the intrinsic scatter in the relation we subtract off in quadrature the measurement uncertainty. The intrinsic scatter in the \textit{M/L} ratio vs. log Mass for the $\mathrm{z\approx0.7}$ and $\mathrm{z\approx1}$ passive field galaxies is $\mathrm{0.18\pm0.05}$ and $\mathrm{0.17\pm0.04}$, respectively. The intrinsic scatter in the \textit{M/L} ratio vs. log Mass for the RXJ1226.9+3332 and RXJ0152.7-1357 cluster galaxies treated as one sample is $\mathrm{0.08\pm0.01}$. This may indicate that the passive field galaxies have less homogeneous stellar populations than the high redshift cluster galaxies. 

\subsection{Line Indices vs. the Velocity Dispersions}
Figure \ref{lineindices} shows the line indices vs. the velocity dispersions of the galaxies.

The $\mathrm{(H\delta_A + H\gamma_A)'}$ lines are stronger for the passive field galaxies than for cluster galaxies at similar redshifts. This indicates that the passive field galaxies have younger stellar populations than the cluster galaxies. 




To find the predicted formation redshift of the passive field galaxies, we use the equation $\mathrm{\Delta (H\delta_A+H\gamma_A)' = -0.126 \Delta log\ age}$ \citep{Jorgensen2013} based on the SSP models in \cite{Thomas2011}. The $\mathrm{\chi^2}$ fits to the zero point offsets of the $\mathrm{(H\delta_A+H\gamma_A)'}$ index imply a formation redshift of $\mathrm{z_{form}=1.40^{+0.60}_{-0.18}}$ for the passive field galaxies. This formation redshift is consistent with that found from the \textit{M/L} ratios. The formation redshift was also determined for the median zero point differences in the $\mathrm{z\approx 0.7}$ and $\mathrm{z\approx 1}$ passive field galaxy subsamples, see Table \ref{tab-formationredshift}. The $\mathrm{H\zeta_A}$ lines for the passive field galaxies indicate a formation redshift consistent with those found from the higher order Balmer lines but due to the large uncertainties on this index for the Perseus and A194 galaxies, more accurate measurements need to be taken for these measurements to provide additional constraints. The zero point differences for the C4668 and Fe4383 lines imply a formation redshift of $\mathrm{z_{form}\approx 1.3-1.4}$, generally consistent with the higher order Balmer lines.

To calculate the intrinsic scatter in the relation for Fe4383, we subtract off in quadrature the measurement uncertainty. We find an intrinsic scatter for Fe4383 of $\mathrm{0.16\pm0.05}$ and $\mathrm{0.26\pm0.08}$ for the $\mathrm{z\approx 0.7}$ and $\mathrm{z\approx 1}$ passive field galaxy samples, respectively. The intrinsic scatter for Fe4383 of the local comparison sample is $\mathrm{0.05\pm0.01}$ and for RXJ1226.9+3332 is $\mathrm{0.04\pm0.01}$. Therefore, the $\mathrm{z\approx 0.7}$ and $\mathrm{z\approx 1}$ passive field galaxies show 4 and 6.5 times more intrinsic scatter, respectively, than that found for the cluster galaxies at similar redshifts. This supports the results from the scatter in the \textit{M/L}-mass relation, further indicating that the passive field galaxies have less homogeneous stellar populations than the cluster galaxies at similar redshifts.

\subsection{Index-Index Plots and Comparison to Stellar Population Models}
Figure \ref{SSPmodels} shows the line indices versus each other for the field galaxies together with the local cluster sample and the RXJ1226.9+3332 cluster sample. Panel (a) shows SSP models from \cite{Thomas2011} with [$\mathrm{\alpha /Fe}$]=0, 0.3, 0.5 and [M/H]=-0.33, 0.00, 0.35, 0.67. Panel (b) shows SSP models for $\mathrm{H\zeta_A}$ based on \cite{MarastonandStromback2011} SEDs for $\mathrm{[\alpha/Fe]=0}$, while CN3883 is based on $\mathrm{CN_2}$ from \cite{Thomas2011}. The models cover [M/H] from -0.33 to 0.67 and ages of 1-15 Gyr. Panel (c) shows SSP models from \cite{Thomas2011} for [$\mathrm{\alpha /Fe}$]=0.3 with metallicities [M/H] between -0.33 to 0.67 and ages of 1-15 Gyr. The figure shows that the sample of field galaxies contains a range of abundance ratios [$\mathrm{\alpha/Fe}$], see panel (a), and young stellar populations (1-5 Gyr), see panels (b) and (c). Compared to the field galaxies, the cluster galaxies span a similar range in the line indices and therefore have a similar range in [M/H] and [$\mathrm{\alpha /Fe}$].




\section{Discussion} \label{discussion}
Here we discuss the results of this work in the broader context of the published literature. The following topics will be discussed: \S \ref{sizediscussion} size and velocity dispersion evolution, \S \ref{fpdiscussion} the FP, and \S \ref{agesdiscussion} the stellar populations and ages.

\begin{figure*}
	\epsscale{1.2}
	\plotone{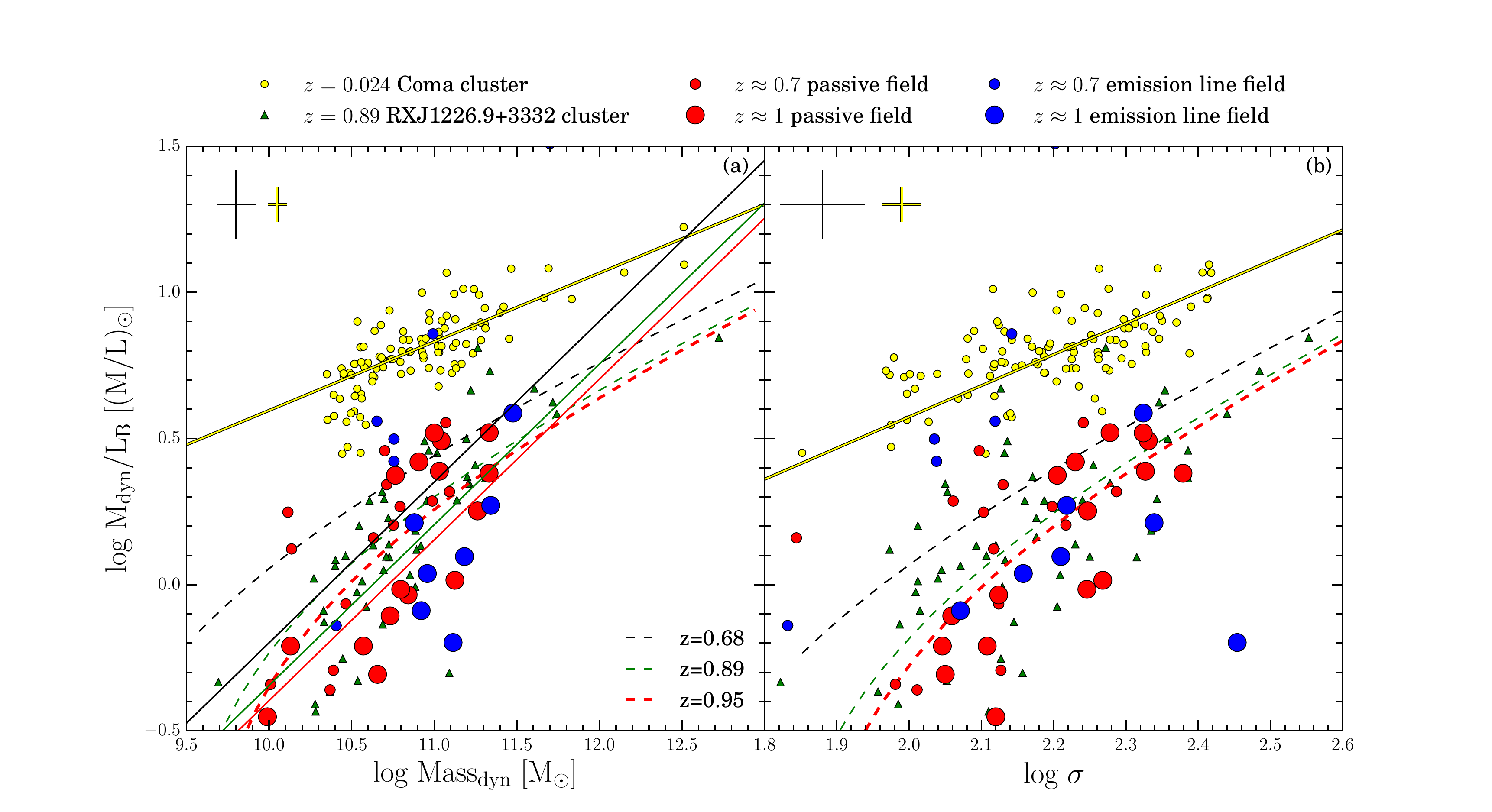}
	\caption{(a) \textit{M/L} ratio vs. log $\mathrm{Mass_{dyn}}$. (b) \textit{M/L} ratio vs. log $\sigma$. Symbols as on Figure \ref{lrelsigmalmass}. Solid green line: \textit{M/L} vs. Mass relation for RXJ1226.9+3332 and RXJ0152.7-1357 treated as one sample. Black and red lines: best fit relations for the two high redshift clusters offset to the median zero point of the $\mathrm{z\approx 0.7}$ and $\mathrm{z\approx 1}$ passive field galaxies, respectively. Dashed lines are the predictions based on models from \citet{Thomas2005} for the relations of the median redshifts of the two passive field galaxy samples ($\mathrm{z=0.68; z=0.95}$) and of the RXJ1226.9+3332 cluster (z=0.89). \label{lmllmasslsigma}}
\end{figure*}

\subsection{Size and Velocity Dispersion Evolution}\label{sizediscussion}
Our sample of passive field galaxies follows the same size-mass and velocity dispersion-mass relations as found for the Coma cluster sample and we therefore conclude that at a given dynamical mass, the passive field galaxies show no significant evolution of size or velocity dispersion between $\mathrm{z \approx 1}$ and the present. \cite{Oldham2017} find an extremely small amount of structural evolution in their sample of z=0.545 cluster galaxies. In contrast to this, several studies have found that galaxies at a given mass are smaller at higher redshifts \citep{Trujillo2007, vanDokkum2010, Newman2012, Belli2014}. However, these authors all use stellar masses while we use dynamical masses in our relations and models. It is unclear if the different method of mass estimation is the reason for the disagreement between our results.

\begin{figure*}
	\epsscale{1.1}
	\plotone{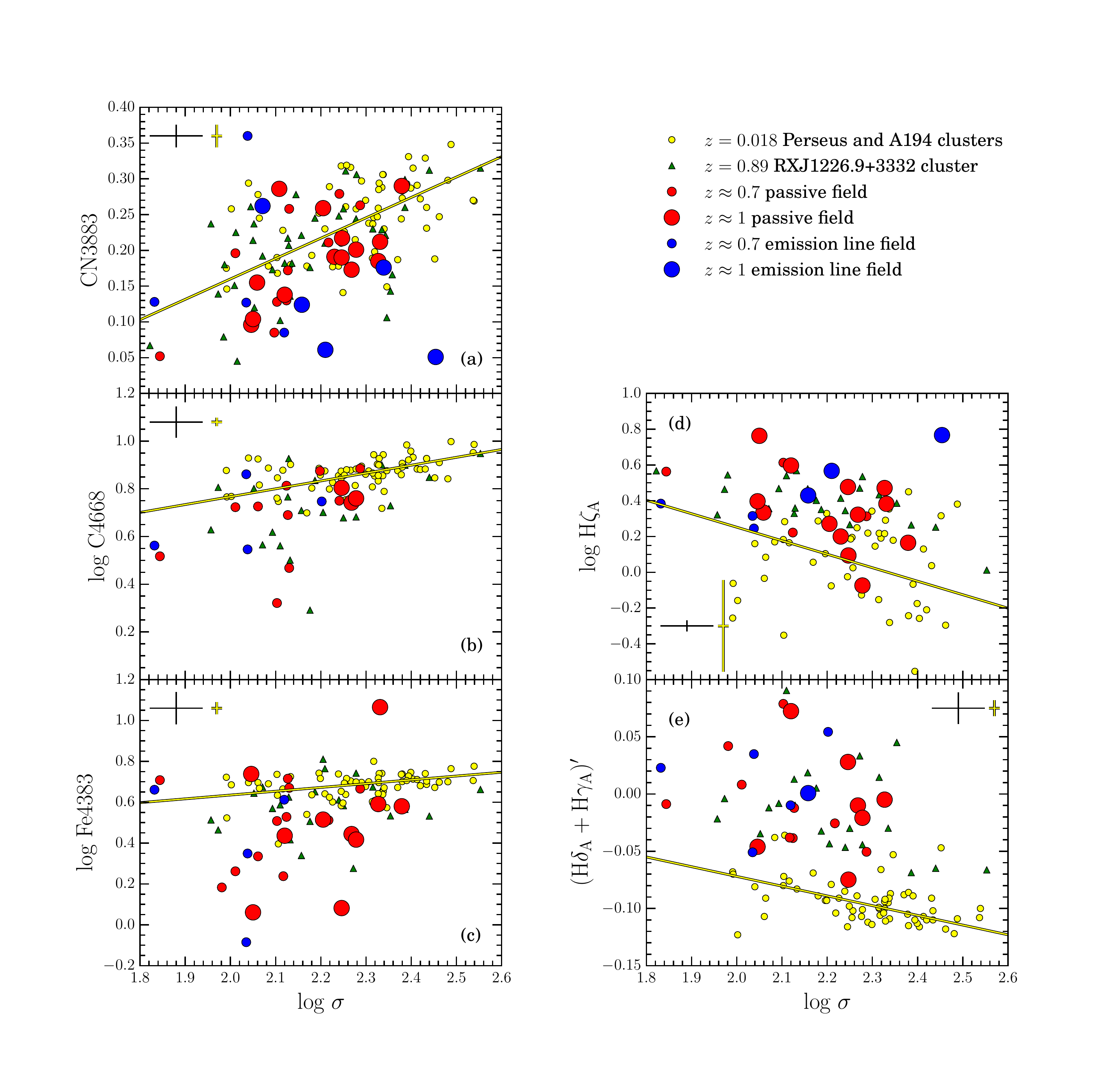}
	\caption{Line indices in the blue vs. velocity dispersions. Symbols are the same as on Figure \ref{lrelsigmalmass}. The Balmer lines are stronger and the metal lines are weaker for the field galaxies when compared to the local sample, as expected under the assumption of passive evolution. \label{lineindices}}
\end{figure*}

\cite{Jorgensen2013}, \cite{Jorgensen2014}, and \cite{Saglia2010} use dynamical masses for deriving the size-mass and velocity dispersion-mass relations. \cite{Jorgensen2013} and \cite{Jorgensen2014} found no significant evolution in galaxy sizes or velocity dispersions as a function of redshift at a given dynamical mass for their sample of $\mathrm{z=0.86}$ and $\mathrm{z=1.27}$ cluster galaxies, respectively. \cite{Saglia2010} analyze field galaxies and galaxies in less rich cluster environments (compared to \citealt{Jorgensen2013}), and found that both field and cluster galaxies at a given mass are smaller and increase in velocity dispersion at higher redshifts. \cite{Jorgensen2013} already established that the differences between our conclusions are not due to selection effects, as \cite{Saglia2010} use the same limit in $\mathrm{r_e}$ as ours, excluding galaxies with $\mathrm{r_e<1\ kpc}$ from the samples. For a better comparison with our results, more studies of size evolution using dynamical masses are needed.
 
Our sample of passive field galaxies follow the same size-mass and velocity dispersion-mass relations as the cluster galaxy comparison sample at similar redshift. The \cite{Saracco2017} sample of passive field galaxies at $\mathrm{z=1.3}$ also follows the same size-mass scaling relations as their sample of cluster galaxies, meaning that at fixed stellar mass, they have the same structural parameters. They also find a significant lack of passive bulge-dominated field galaxies at $\mathrm{M_{dyn}>2\cdot10^{11} M_\odot}$ and $\mathrm{r_e>5 kpc}$. Our sample also displays these trends but this may be due to selection effects since our sample is by no means complete. 

\begin{figure*}
	\epsscale{1.2}
	\plotone{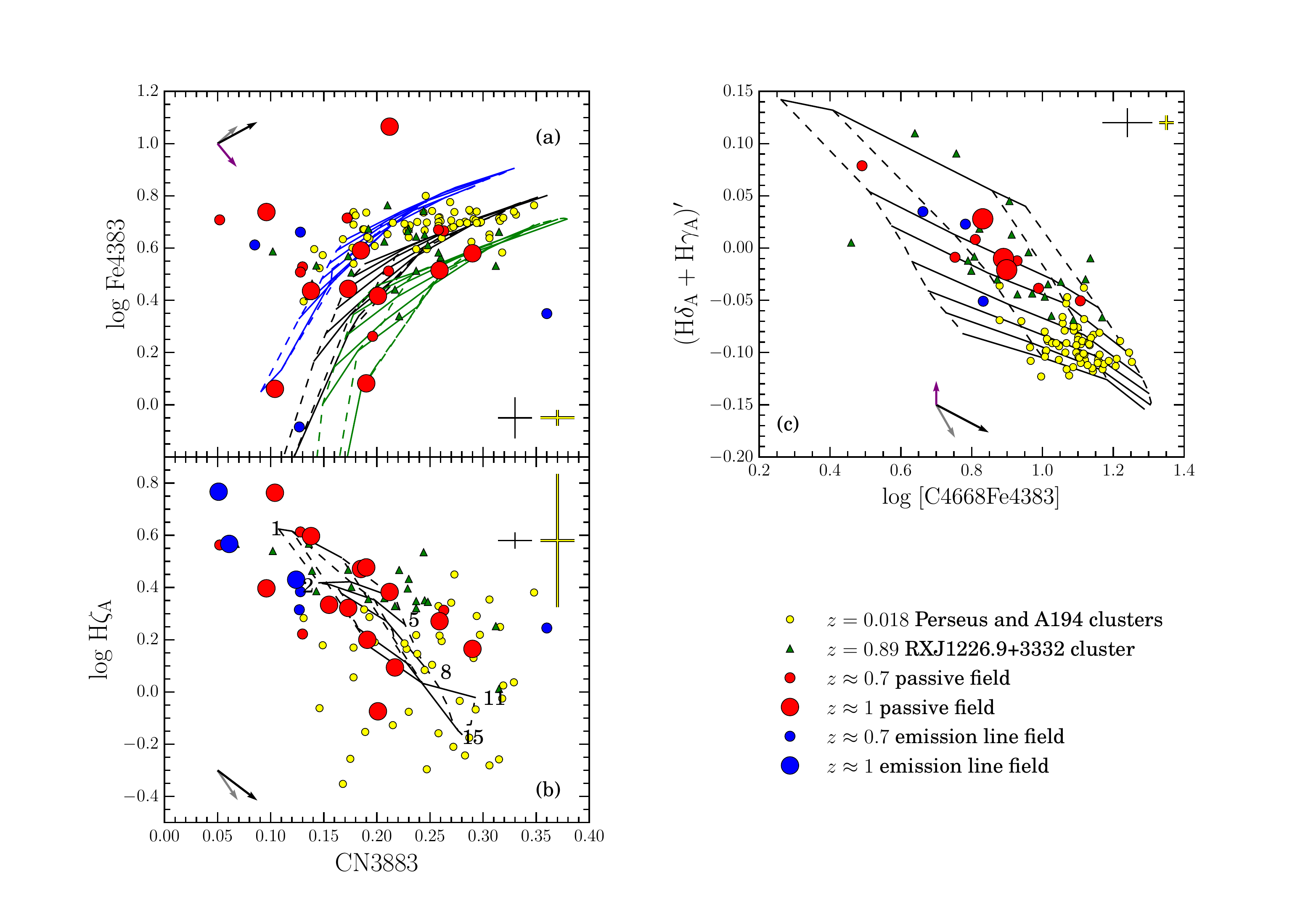}
	\caption{Panel (a) shows SSP models from \cite{Thomas2011} with [$\mathrm{\alpha/Fe]=0}$ in blue, 0.3 in black and 0.5 in green and [M/H]=-0.33, 0.00, 0.35, 0.67 and ages of 1-15 Gyr. Panel (b) shows $\mathrm{H\zeta_A}$ models based on \citet{MarastonandStromback2011} and CN3883 from $\mathrm{CN_2}$ from \cite{Thomas2011} models. The models are shown for [$\mathrm{\alpha/Fe]=0}$ and [M/H]=-0.3, 0.0, and 0.3. The locations of ages 1, 2, 5, 8, 11 and 15 Gyr for solar metallicity are labeled. Panel (c) shows SSP models from \cite{Thomas2011} for [$\mathrm{\alpha/Fe]=0.3}$ with metallicities [M/H] between -0.33 to 0.67 and ages of 1-15 Gyr. The arrows show the average model predictions for $\mathrm{\Delta\ log\ age = 0.3\ (grey), \Delta [M/H] = 0.3\ (black), and\ \Delta [\alpha/Fe] = 0.3\ (purple)}$. These figures confirm that the field galaxies contain a range of abundance ratios [$\mathrm{\alpha/Fe}$], panel (a), and young stellar populations (1-5 Gyr), panels (b) and (c).\label{SSPmodels} }\end{figure*}

\subsection{The Fundamental Plane}\label{fpdiscussion}

We find that the slope of the \textit{M/L} ratio vs. dynamical mass of our sample of $\mathrm{z\approx0.7}$ and $\mathrm{z\approx1}$ early-type field galaxies is steeper than that of Coma. Supporting this conclusion, steeper FP slopes at high redshift have also been reported by \cite{diSeregoAlighieri2005}, \cite{Treu2005b, Treu2005}, and \cite{vanderWel2005} for their samples of field galaxies. This effect is also seen in high-density environments as shown by \cite{Jorgensen2006, Jorgensen2007} and \cite{Jorgensen2013}. However, \cite{Gebhardt2003} find no significant difference between the slopes of the FP for their sample of field galaxies and the local comparison sample. The discrepancy between our results and those of \cite{Gebhardt2003} can possibly be attributed to their sample not containing galaxies of low enough masses, with only two galaxies in the redshift range $\mathrm{0.75<z<1.0}$ with masses $\mathrm{< 10^{10.8} M_{\odot}}$.
 
 The steepening of the FP slope is consistent with ``downsizing" \citep{Cowie1996}, where the high mass galaxies formed their stars at higher redshift and in a shorter time period than low mass galaxies. Figure \ref{lmllmasslsigma} shows the predicted locations of the relations for passive evolution models from \cite{Thomas2005} for the median redshifts of the $\mathrm{z\approx0.7}$ and $\mathrm{z\approx1}$ passive field galaxies and the RXJ1226.9+3332 ($\mathrm{z=0.89}$) cluster galaxies showing galaxies with high mass forming their stars at high redshift and low mass galaxies forming their stars more recently. All three model lines use the high density environment predictions. The data roughly follow the models, implying that the passive evolution model and ``downsizing" can explain our results. However, the passive field galaxies appear to show more evolution in the \textit{M/L} ratio vs. log $\mathrm{\sigma}$ relation than predicted by the models, meaning that age depends more strongly on the velocity dispersion for this sample than stated in \cite{Thomas2005}.
 
\begin{figure*}
  	\epsscale{1.2}
  	\plotone{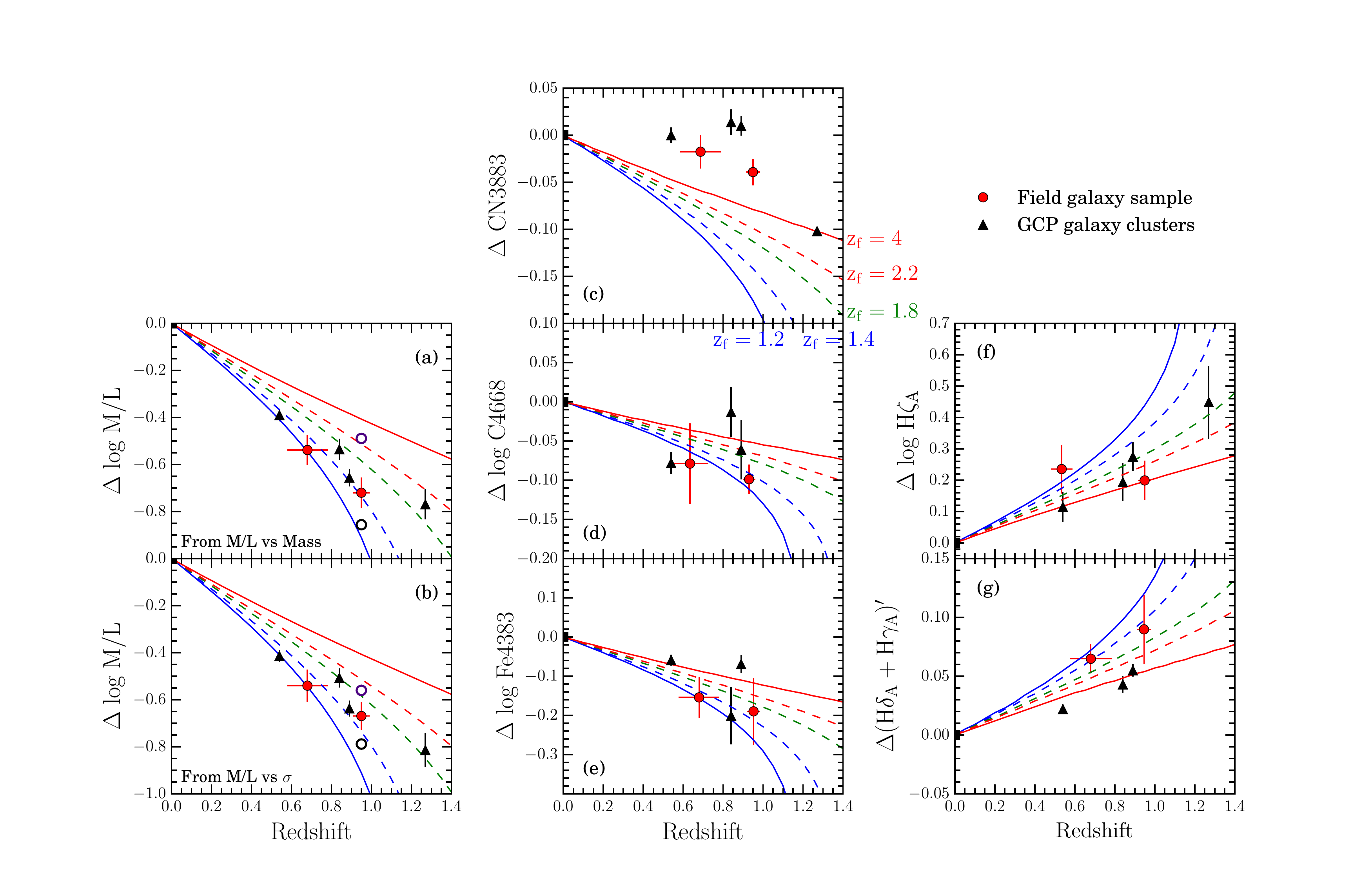}
  	\caption{Zero point differences of the scaling relations as a function of redshift. The zero point differences are derived as $\mathrm{\Delta zp = zp_{FG} - zp_{low-z}}$. The random uncertainties in the zero point differences are shown as error bars. Red circles: $\mathrm{z\approx 0.7}$ and $\mathrm{z\approx 1}$ passive field galaxy samples. Black triangles: galaxy clusters from the GCP. Black and purple circles: low-mass and high-mass passive field galaxy subsamples, respectively, for the $\mathrm{z\approx 1}$ sample. The lines overplotted show passive evolution models for $\mathrm{z_{form}}$ as labeled based on SSP models from \citet{MarastonandStromback2011} for $\mathrm{H\zeta_A}$, from \citet{Maraston2005} for the \textit{M/L} ratios and from \citet{Thomas2011} for the remainder of the indices. The zero offset in these models was set to z=0, while the zero offset in the calculations for Table \ref{tab-formationredshift} was set to z=0.024 for the plots that use Coma as the local comparison and z=0.018 for the plots that use Perseus. The median redshift of the field galaxies varies between panels because not all of the field galaxies have measurements for all line indices. The results for massive clusters are from the GCP \citep{Jorgensen2013, Jorgensen2014}. These figures show that the passive field galaxies are consistent with the passive evolution model. The zero point differences in the \textit{M/L} ratios agree with the formation redshift from most of the line indices. The models are not consistent with our data for CN3883. \label{formationredshift}}
  \end{figure*}

For a deeper understanding of the evolution of field galaxies as a function of mass, we used the zero point differences of the \textit{M/L} ratios to determine formation redshifts for the low-mass and high-mass passive field galaxy subsamples, see Table \ref{tab-formationredshift} and Figure \ref{formationredshift}. These findings are consistent with those from other authors. For example, \cite{vanDokkum2007} find $\mathrm{z_{form}=1.95^{+0.10}_{-0.08}}$ for field galaxies with $\mathrm{M_{dyn} >10^{11}M_{\odot}}$ while for cluster galaxies within the same mass range they find $\mathrm{z_{form}=2.01^{+0.22}_{-0.17}}$ (see also \citealt{vanderWel2005}, \citealt{Oldham2017}). Similarly, \cite{Treu2005} found $\mathrm{z_{form}\approx 1.2}$ and $\mathrm{z_{form}>2}$ for their low mass and high mass subsamples, respectively. They interpret the results to mean that the majority of the stellar mass in all systems was formed at $\mathrm{z>2}$, but that the low mass systems had secondary episodes of activity that revived the older population. 

\begin{figure}[h!]
  	\epsscale{1.25}
  	\plotone{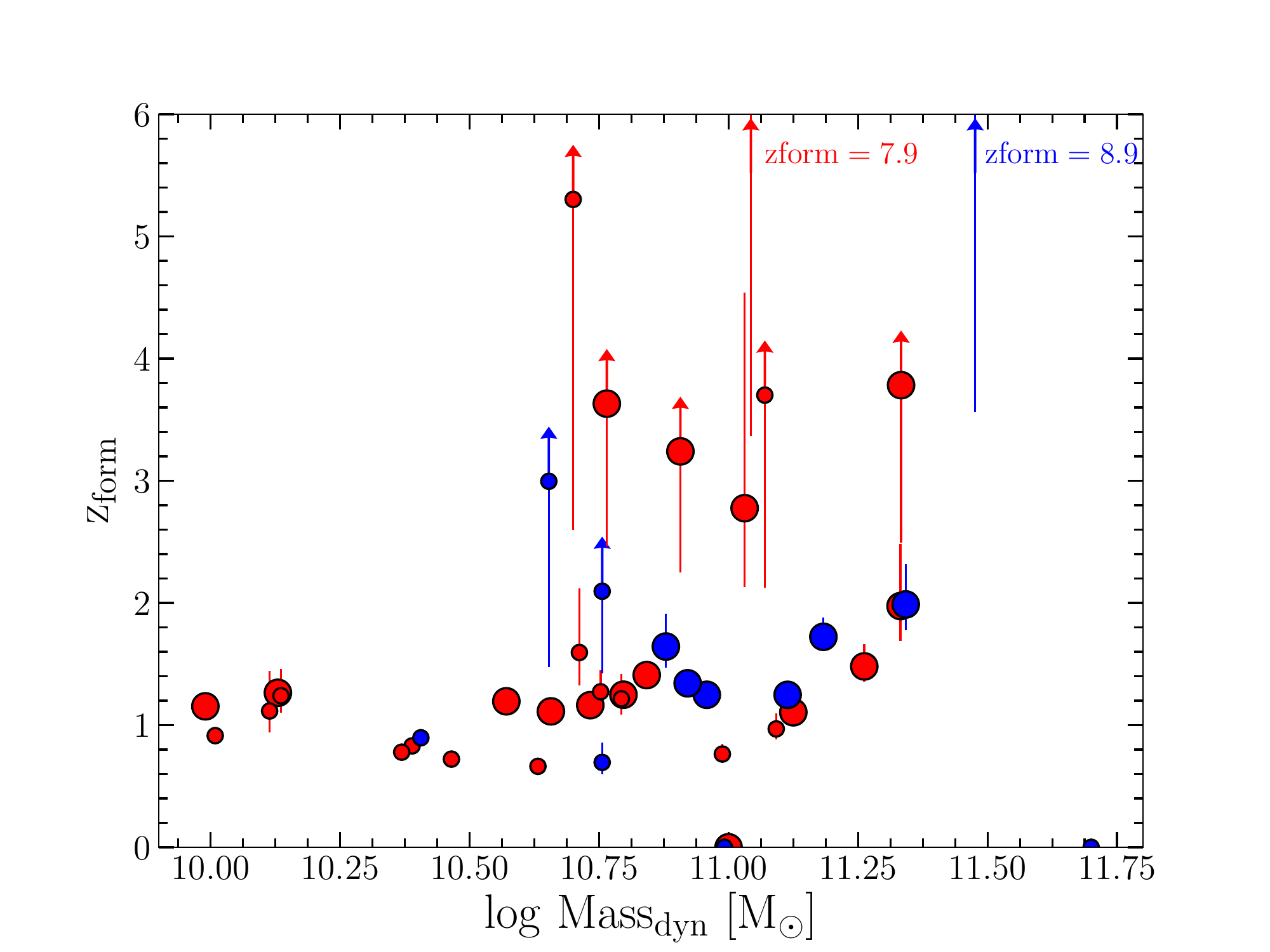}
  	\caption{Formation redshifts for the individual field galaxies as a function of dynamical mass. Small red circles: $\mathrm{z\approx0.7}$ passive field galaxies. Large red circles: $\mathrm{z\approx1}$ passive field galaxies. Small blue circles: $\mathrm{z\approx0.7}$ emission line field galaxies. Large blue circles: $\mathrm{z\approx1}$ emission line field galaxies. Arrows indicate uncertainties on $\mathrm{z_{form}}$ that extend to the Big Bang. Three galaxies have formation redshifts calculated to be negative. Since this is unphysical, their formation redshifts are set to 0. The two galaxies with $\mathrm{z_{form}>6}$ are from the $\mathrm{z\approx 1}$ sample. \label{zformmasssigma}}
  	\end{figure}

We investigate this further by showing the formation redshifts of individual field galaxies derived from the zero point differences in the \textit{M/L} ratios versus dynamical masses, see Figure \ref{zformmasssigma}. All of the field galaxies with log $\mathrm{Mass_{dyn}\lesssim10.6\ M_{\odot}}$ have experienced recent star formation. Above this mass, our sample contains a mix of galaxies, some that have experienced recent star formation and others that contain only older stellar populations. Since our sample is not complete, this analysis likely contains selection effects. A larger and more complete sample is required to determine if this is a general feature of passively evolving field galaxies.
	
 
 In the case of cluster galaxies, \cite{Jorgensen2013} found $\mathrm{z_{form}=1.24\pm0.05}$ for low mass galaxies and $\mathrm{z_{form}=1.95^{+0.30}_{-0.20}}$ for high mass galaxies. The consistency shown between the field and cluster galaxy populations in terms of formation epoch strongly suggests that galaxy age depends more on mass, rather than environment. \cite{Rettura2011} suggest that the timescale of the star formation histories of galaxies is determined by environment, but the timing of galaxy formation is regulated by the galaxy mass. This means that at a given stellar mass, field and cluster galaxies have roughly the same age (or $\mathrm{z_{form}}$) but galaxies in clusters tend to have a less extended phase of star formation. More extended star formation activity should lead to lower $\mathrm{[\alpha/Fe]}$ abundances as the products of Type Ia supernovae are recycled with those from core-collapse supernovae. The $\mathrm{\alpha}$-element abundances, as derived from line index fitting, may provide an independent assessment of the length of the star formation in galaxies as a function of environment. However, the typical star formation timescale difference between cluster and field reported in \cite{Rettura2011} is $<1$ Gyr. This is too short of a difference to be detected in $\mathrm{\alpha}$-element abundances.
 
 \begin{deluxetable*}{lllll}
 	
 	
 	
 	
 	\tablecaption{Formation Redshift Results from SSP Models\label{tab-formationredshift}}
 	

 	\tablehead{\multicolumn{1}{c}{\multirow{2}[3]{*}{Relation}} & \multicolumn{3}{c}{Field Galaxies} & \colhead{Galaxy Clusters} \\
 		\cmidrule(lr){2-4} \cmidrule(lr){5-5} 
 		& \colhead{$\mathrm{\chi^2}$ Fit} & \colhead{$\mathrm{z\approx0.7}$} & \colhead{$\mathrm{z\approx1}$} & \colhead{z=0.86} \\ 
 		\colhead{(1)} & \colhead{(2)} & \colhead{(3)} & \colhead{(4)} & \colhead{(5)} } 
 	
 	\startdata
 	$\mathrm{\Delta log\ \textit{M/L} = 0.935 \Delta log\ age}$ & $1.35^{+0.10}_{-0.07}$ & $1.16^{+0.17}_{-0.11}$ & $1.39^{+0.12}_{-0.09}$ & \nodata \\
 	\hspace{3mm} log \textit{M/L} (low-mass) & \nodata & \nodata & $1.24^{+0.10}_{-0.07}$ & $1.24\pm0.05$ \\
 	\hspace{3mm} log \textit{M/L} (high-mass) & \nodata & \nodata & $2.16^{+0.81}_{-0.39}$ & $1.95^{+0.3}_{-0.2}$ \\
 	$\mathrm{\Delta(H\delta_A + H\gamma_A)'=-0.126\Delta log\ age}$ & $1.40^{+0.60}_{-0.18}$ & $1.32^{+0.47}_{-0.22}$ & $1.49^{+1.22}_{-0.29}$ & $>2.8$ \\
 	$\mathrm{\Delta log\ H\zeta_A=-0.456 \Delta log\ age}$ & $2.60^{+\infty}_{-0.62}$ & $0.94^{+0.62}_{-0.19}$ & $3.50^{+\infty}_{-1.57}$ & \nodata \\
 	$\mathrm{\Delta log C4668=0.121 \Delta log\ age}$ & $1.29^{+\infty}_{-\infty}$ & $0.96^{+\infty}_{-\infty}$ & $1.29^{+0.25}_{-0.13}$ & \nodata \\
 	$\mathrm{\Delta log Fe4383=0.272 \Delta log\ age}$ & $1.38^{+7.75}_{-0.24}$ & $1.19^{+1.81}_{-0.29}$ & $1.53^{+20.1}_{-0.39}$ & \nodata \\
 	\enddata
 	\tablecomments{Column 1: Scaling relation. Column 2: Formation redshift determined with $\mathrm{\chi^2}$ fits to the zero point offsets in the scaling relation for the passive field galaxies. The formal uncertainties derived from the fits are unrealistically low, so the uncertainties listed were derived as the random uncertainties on the median zero points. Column 3-4: Formation redshift determined from the median zero point offsets in the scaling relation for the $\mathrm{z\approx 0.7}$ and $\mathrm{z\approx 1}$ passive field galaxy subsamples, respectively. Column 5: Formation redshift determined from the zero point offsets in the scaling relation for the galaxy clusters at z=0.86 \citep{Jorgensen2013}}
 	
 	
 	
 \end{deluxetable*}
 
 The intrinsic scatter in the \textit{M/L} vs. Mass relation for Coma is consistent with that of the z=0.86 clusters RXJ1226.9+3332 and RXJ0152.7-1357 treated as one sample. In contrast, \cite{Treu2005} and \cite{vanderWel2005} find that the intrinsic scatter in the FP decreases with decreasing redshift, after allowing for evolution of the slopes. 
 
 The intrinsic scatter in the \textit{M/L} vs. Mass relation for the passive field galaxies is $\mathrm{\sim2.5}$ higher than the intrinsic scatter of the cluster galaxies at similar redshifts. This may indicate that the scatter in the \textit{M/L} vs. mass relation is dependent on environment and not redshift. However, we recognize that this result depends on how precisely we understand and can quantify our measurement error. In contrast to this, \cite{Bernardi2006} show that the scatter in the FP at $0.05<z<0.14$ does not depend on environment.
\subsection{Stellar Populations and Ages}\label{agesdiscussion}

Line indices are used to determine characteristics of stellar populations such as ages and metallicities. For the cluster galaxies in \cite{Jorgensen2013}, the offsets in the higher order Balmer lines imply older ages than those found from the offsets in the \textit{M/L} ratios. In contrast to this, for our sample of passive field galaxies, the ages determined from the strength of the higher order Balmer lines agree with the formation redshifts determined from the zero point differences in their \textit{M/L} ratios. We confirmed there was no bias in our sample by calculating the formation redshift for only those field galaxies that had measured higher order Balmer lines. We then compared that result to the formation redshift determined from the entire sample of passive field galaxies and found no significant difference. Consistent with our results, \cite{Schiavon2006} find, for red field galaxies at $\mathrm{z\sim0.9}$, a formation redshift of $\mathrm{z_{form}\approx1.1-1.3}$ when modeled using SSP models with supersolar metallicity.

One line index we use in our analysis is CN3883. Our sample of passive field galaxies exhibit weaker CN3883 lines than the cluster galaxies at similar redshifts. However, because there are no models published for CN3883, the predictions for this index were based on $\mathrm{CN_2}$ models and the empirical relation between CN3883 and $\mathrm{CN_2}$. The passive evolution models predict a weakening of CN3883 that we do not detect in our data. We suggest that better or updated models are needed for this index. 

Our analysis of the zero point differences of the line indices and \textit{M/L} ratios of the passive field galaxies is consistent with the passive evolution model. In agreement with this, \cite{Gallazzi2014} and \cite{Choi2014} find that their samples of $\mathrm{z\sim0.7}$ quiescent field galaxies are consistent with passive evolution. These authors also find no significant evolution in metallicities since $\mathrm{z\sim0.7}$.

An important line index in our analysis centers on the Fe4383 lines because we find that the $\mathrm{z\approx0.7}$ and $\mathrm{z\approx1}$ passive field galaxies show 4 and 6.5 times more intrinsic scatter, respectively, than the cluster galaxies at similar redshifts. This may indicate that the passive field galaxies are more diverse in Fe4383 than the cluster galaxies. One potential explanation for this is that, as compared to the cluster galaxies, the passive field galaxies have experienced a larger range in the duration of their star formation episodes, consistent with \cite{Rettura2011}.

\section{Summary and Conclusions} \label{conclusions}
To complement the study of cluster galaxies in \cite{Jorgensen2013}, we carried out an analysis of field galaxies observed as part of the GCP \citep{Jorgensen2005}. From a total of 82 galaxies, we selected a sample of 43 bulge-dominated field galaxies using the following criteria: $\mathrm{S/N \geq 20\ per\ \AA}$ in the restframe, $\mathrm{z\geq0.3}$, and $\mathrm{n_{ser}\geq 1.5}$. We use effective radii and surface brightnesses published in \cite{Chiboucas2009} and \cite{Jorgensen2013}. We use GMOS-N spectroscopy to measure velocity dispersions and line index strengths, published for the first time in this paper. We divide our sample of field galaxies into two redshift bins: $\mathrm{0.3<z<0.8}$ and $\mathrm{0.8<z<1.2}$ with median redshifts of $\mathrm{z\approx0.7}$ and $\mathrm{z\approx1}$, respectively. We separated the sample of the 43 bulge-dominated field galaxies into passive and emission-line objects and used the 30 passive galaxies to conduct an analysis on the evolution of their FP and their stellar populations through line indices. Line indices were further modeled using the SSP models of \cite{Thomas2011} and \cite{MarastonandStromback2011} in order to establish formation redshifts. Our main results are as follows:

 \begin{enumerate}

\item At a given dynamical mass, the passive field galaxies show no significant evolution in size or velocity dispersion between $\mathrm{z\approx1}$ and the present, when compared to local cluster counterparts.

\item The Fundamental Plane and \textit{M/L} ratios vs. dynamical masses and velocity dispersions show that the passive field galaxies follow relations with steeper slopes than the Coma cluster relations. The steeper slopes of the passive field galaxies are consistent with the slopes of the cluster galaxies in the GCP at similar redshifts. This shows that the low mass galaxies formed their stars more recently while the high mass galaxies formed their stars at higher redshifts. To further test this ``downsizing" scenario, we divided our $\mathrm{z\approx1}$ passive field galaxy sample into two (low- and high-velocity dispersion) subsamples. Using the zero point offsets of the \textit{M/L} ratio-mass relation for these two subsamples, we find formation redshifts of $\mathrm{z_{form}=1.24_{-0.07}^{+0.10}}$ and $\mathrm{z_{form}=2.15_{-0.38}^{+0.80}}$ for the low- and high-velocity dispersion subsamples, respectively. For the cluster galaxies at similar redshifts, the same relation implies formation redshifts of $\mathrm{z_{form}=1.24\pm0.05}$ and $\mathrm{z_{form}=1.95_{-0.2}^{+0.3}}$ for the low- and high-velocity dispersion subsamples, respectively. This suggests that the ages of the galaxies depend more on mass than on environment. From the formation redshifts of individual galaxies, we determine that all of the field galaxies with log $\mathrm{Mass_{dyn}\lesssim10.6\ M_{\odot}}$ have experienced recent star formation. Above this mass, our sample contains a mix of galaxies, some that have experienced recent star formation and others that contain only older stellar populations.

\item The \textit{M/L} ratios give a formation redshift $\mathrm{z_{form}=1.35_{-0.07}^{+0.10}}$ for the passive field galaxy samples. Analysis of the higher order Balmer lines provide formation redshifts of $\mathrm{z_{form}=1.40^{+0.60}_{-0.18}}$ for the passive field galaxies, consistent with those derived from the \textit{M/L} ratios. This is also supported by the C4668 and Fe4383 indices, which indicate formation redshifts of $\mathrm{z_{form}\approx1.3-1.4}$. We note that the models fail to reproduce the very small change in CN3883 index values with redshift, which suggests that a revision of the models is needed when trying to reproduce this feature. For the cluster galaxies, the higher order Balmer lines imply formation redshifts of $\mathrm{z_{form}> 2.8}$. This indicates that the passive field galaxies contain younger stellar populations than the cluster galaxies at similar redshifts, in disagreement with consistent ages derived between the two samples from the offsets in the \textit{M/L}-mass relation.


\end{enumerate}

The comparison between our sample of passive field galaxies and cluster members may indicate that environment causes differences in the stellar populations of bulge-dominated galaxies. Namely, the increased scatter in the \textit{M/L}-mass relation and Fe4383 lines for the passive field galaxies implies that the field galaxies are less homogeneous than cluster galaxies at similar redshifts. However, we are aware of the limitations of our small sample size and suggest that a larger, systematically selected sample is needed to confirm the possible environmentally-driven differences in galaxy stellar populations at intermediate redshifts.

\acknowledgments
The editor, Christopher Conselice, and anonymous referee are thanked for their useful comments which have improved this paper. C.W. and R.S.F. acknowledge support provided by the Oregon NASA Space Grant Consortium. R.D. gratefully acknowledges the support provided by the BASAL Center for Astrophysics and Associated Technologies (CATA), and by FONDECYT grant No. 1130528. This research made use of Astropy, a community-developed core Python package for Astronomy (Astropy Collaboration, 2013).

Based on observations obtained at the Gemini Observatory (processed using the Gemini IRAF package), which is operated by the Association of Universities for Research in Astronomy, Inc., under a cooperative agreement with the NSF on behalf of the Gemini partnership: the National Science Foundation (United States), the National Research Council (Canada), CONICYT (Chile), Ministerio de Ciencia, Tecnolog\'{i}a e Innovaci\'{o}n Productiva (Argentina), and Minist\'{e}rio da Ci\^{e}ncia, Tecnologia e Inova\c{c}\~{a}o (Brazil). The data were obtained while also the Science and Technology Facilities Council (United Kingdom), and the Australian Research Council (Australia) contributed to the Gemini Observatory.

The spectroscopic data presented in this paper originate from the following Gemini programs: GN-2002B-DD-4, GN-2002B-Q-29, GN-2003B-DD-4, and GN- 2004A-Q-45. Data for the programs GN-2003A-C-1 and GN-2004A-C-8 were acquired through the Gemini Science Archive. In part, based on observations made with the NASA/ESA \textit{Hubble Space Telescope}, obtained from the data archive at the Space Telescope Science Institute.


\bibliographystyle{aasjournal}

\clearpage

\appendix

\section{}\label{appendixA}
\begin{figure*}
	\epsscale{1.1}
	\plotone{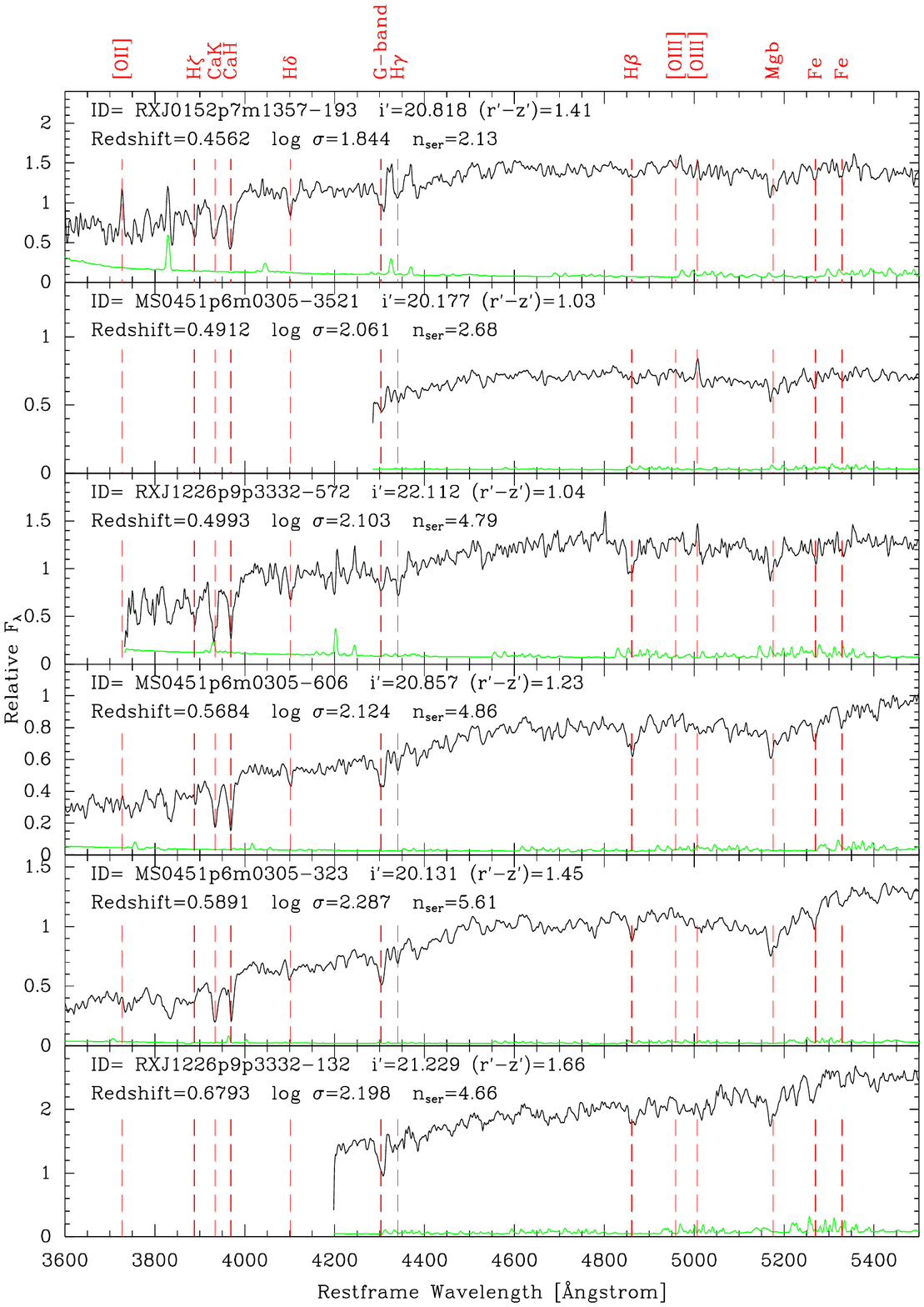}
	\caption{Sample spectra of the field galaxies. Black lines: observed spectra; green lines: four times the random noise in the spectra. Major spectral features are labeled with red dashed vertical lines. This particular figure contains passive bulge-dominated field galaxies by increasing redshift. All field galaxy spectra are available as an electronic figure.\label{spectra}}
\end{figure*}

\begin{figure*}
	\epsscale{1.2}
	\plotone{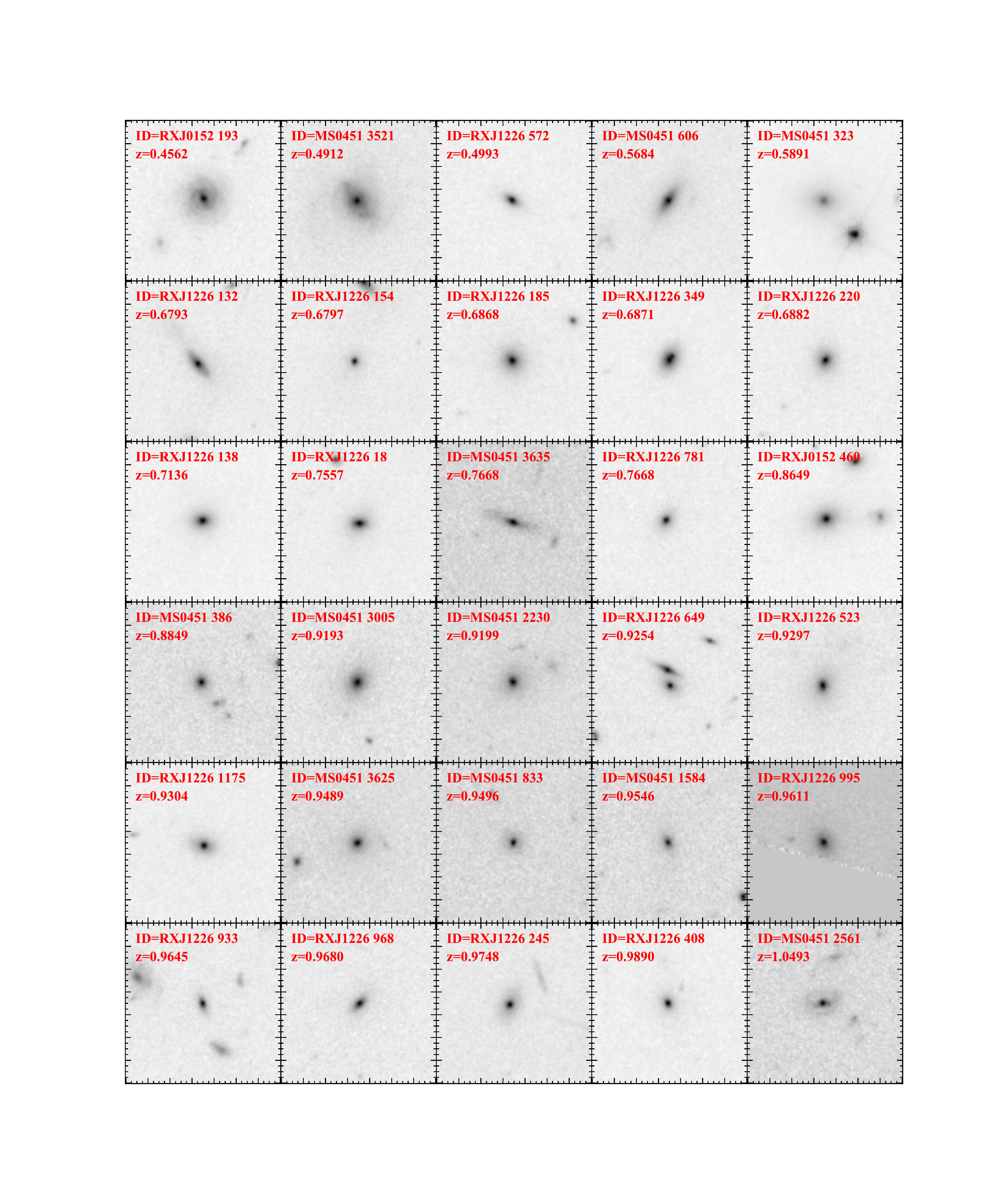}
	\caption{\textit{HST}/ACS images of the passive field galaxies that meet all of the selection criteria in Table \ref{samples}. Each galaxy is centered in its panel and labeled with its ID and redshift. Each panel is 7\arcsec $\times$ 7\arcsec which corresponds to 50 kpc $\times$ 50 kpc at $\mathrm{z\approx0.7}$ and 56 kpc $\times$ 56 kpc at $\mathrm{z\approx1}$.  \label{passivefield}}
\end{figure*}

\begin{figure*}
	\epsscale{1.2}
	\plotone{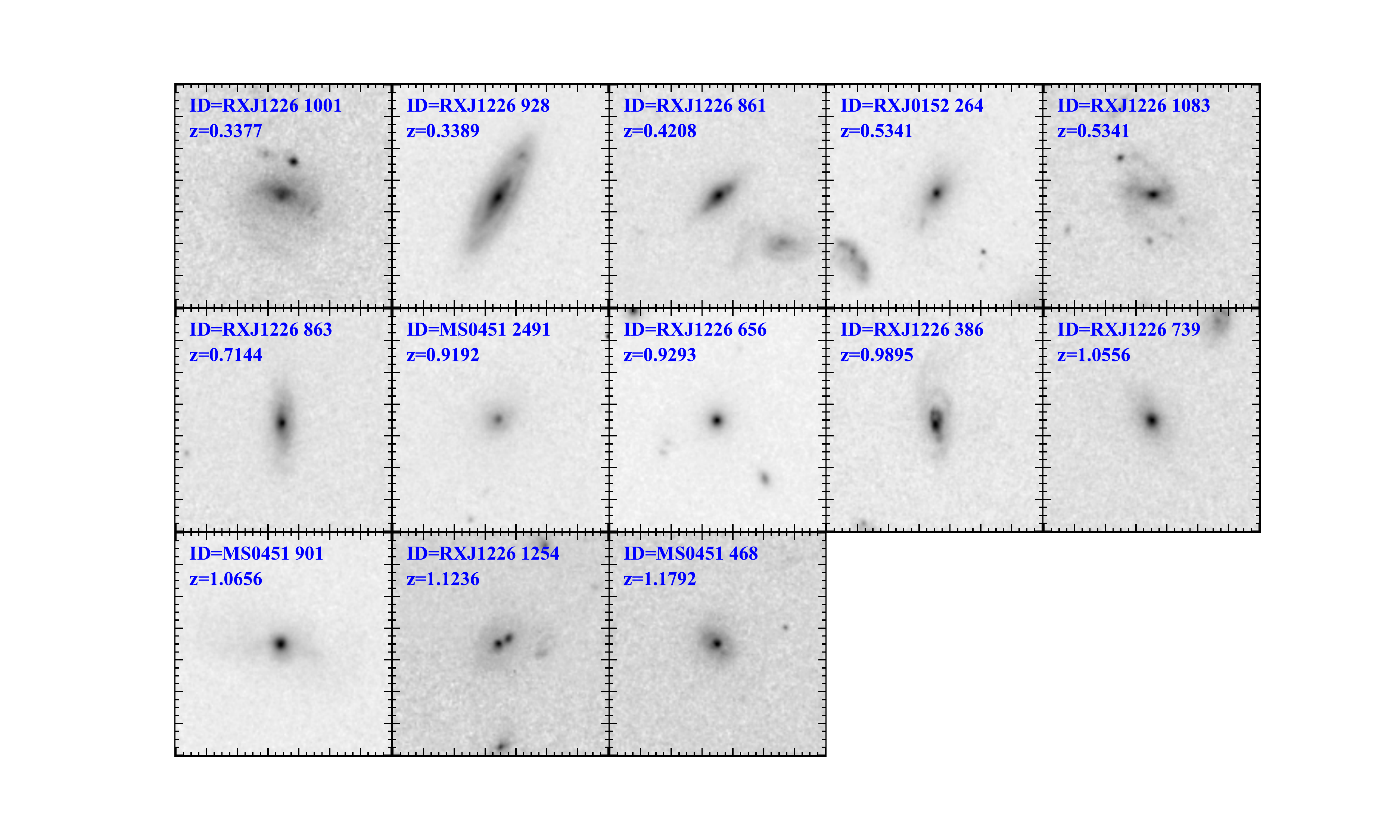}
	\caption{\textit{HST}/ACS images of the emission line field galaxies that meet all of the selection criteria in Table \ref{samples}. Each galaxy is centered in its panel and labeled with its ID and redshift. Each panel is 7\arcsec $\times$ 7\arcsec which corresponds to 50 kpc $\times$ 50 kpc at $\mathrm{z\approx0.7}$ and 56 kpc $\times$ 56 kpc at $\mathrm{z\approx1}$.\label{starformfield}}
\end{figure*}

In this appendix we provide the results from template fitting in Table \ref{tab-speckin} and line indices in Tables \ref{tab-speclineblue} and \ref{tab-speclinevis}. Sample spectra shown in Figure \ref{spectra} with all spectra available in the online journal. Greyscale images of the passive and emission line field galaxies are shown in Figures \ref{passivefield} and \ref{starformfield}. Galaxy IDs are from \cite{Jorgensen2005} and \cite{Jorgensen2013}. The coordinates for the galaxies are published in those papers.

\begin{deluxetable*}{rrrrrrrrrrrr}
	\tablecaption{Results from Template Fitting \label{tab-speckin} }
	\tablewidth{0pt}
	\tabletypesize{\scriptsize}
	\tablehead{
		\colhead{Field/ID} & \colhead{Redshift} & \colhead{$n_{\rm ser}$} & \colhead{Sample\tablenotemark{a}} &
		\colhead{$\log \sigma$} & \colhead{$\log \sigma _{\rm cor}$\tablenotemark{b}} &
		\colhead{$\sigma _{\log \sigma}$} & \multicolumn{3}{c}{Template fractions} & \colhead{$\chi ^2$} & \colhead{S/N\tablenotemark{c}} \\
		& \colhead{}&\colhead{} &\colhead{} &\colhead{} &\colhead{} &\colhead{} & \colhead{B8V} & \colhead{G1V} & \colhead{K0III} & }
	\startdata
	\bf{MS0451.6--0305:} \\
	220& 0.8259&  1.4&    3&  2.192&  2.248&  0.042&   0.42&   0.41&   0.17&    2.7&     45\\
	234& 0.3554&  1.4&    3&  2.121&  2.182&  0.055&   0.68&   0.32&   0.00&    3.2&     75\\
	323& 0.5891&  5.6&    1&  2.237&  2.287&  0.036&   0.13&   0.40&   0.47&    8.5&    114\\
	386& 0.8849&  3.5&    1&  2.171&  2.230&  0.057&   0.00&   0.89&   0.11&    1.7&     15\\
	468& 1.1792&  2.4&    2&  2.147&  2.210&  0.048&   0.46&   0.42&   0.12&    1.4&     25\\
	606& 0.5684&  4.9&    1&  2.041&  2.124&  0.048&   0.14&   0.60&   0.26&    3.8&     71\\
	833& 0.9496&  4.1&    1&  2.141&  2.205&  0.075&   0.05&   0.80&   0.15&    3.2&     16\\
	836& 0.3676&  0.8&    3&  2.072&  2.144&  0.060&   0.95&   0.05&   0.00&    1.4&     44\\
	901& 1.0656&  8.8&    2&  2.398&  2.454&  0.022&   0.80&   0.20&   0.00&    8.5&     95\\
	1584& 0.9546&  4.1&    1&  2.280&  2.331&  0.118&   0.36&   0.20&   0.44&    1.9&     14\\
	2032& 0.5114&  1.1&    3&  2.123&  2.186&  0.043&   0.79&   0.21&   0.00&    2.2&     74\\
	2230& 0.9199&  4.5&    1&  1.939&  2.059&  0.088&   0.00&   0.54&   0.46&    5.7&     25\\
	2491& 0.9192&  2.5&    2&  2.082&  2.158&  0.071&   0.30&   0.46&   0.24&    4.1&     39\\
	2561& 1.0493&  5.0&    1&  2.037&  2.124&  0.154&   0.00&   0.81&   0.19&    3.2&     16\\
	2657& 0.5787&  0.8&    3&  2.033&  2.119&  0.044&   0.31&   0.69&   0.00&    2.7&     47\\
	3005& 0.9193&  2.7&    1&  2.190&  2.247&  0.050&   0.00&   0.69&   0.31&    4.6&     35\\
	3521& 0.4912&  2.7&    1&  1.948&  2.061&  0.062&   0.00&   0.40&   0.60&    2.8&     64\\
	3610& 0.5663&  0.4&    3&  2.038&  2.122&  0.141&   0.77&   0.23&   0.00&    2.8&     32\\
	3625& 0.9489&  5.4&    1&  2.327&  2.379&  0.064&   0.00&   0.52&   0.48&    5.2&     29\\
	3635& 0.7668&  3.8&    1&  1.999&  2.097&  0.063&   0.00&   0.83&   0.17&    1.6&     14\\
	3697& 0.5674&  1.3&    3&  1.963&  2.072&  0.141&   0.65&   0.17&   0.18&    2.5&     23\\
	3792& 0.8996&  0.5&    3&  2.355&  2.407&  0.105&   0.73&   0.27&   0.00&    1.7&     19\\
	\bf{RXJ0152.7--1357:} \\
	103& 0.6406&  \nodata&    7&  1.909&  1.934&  0.053&   0.08&   0.71&   0.20&    0.9&     35\\
	155& 0.9955&  \nodata&    7&  1.418&  1.446&  0.109&   0.51&   0.39&   0.10&    1.7&     22\\
	193& 0.4562&  2.1&    1&  1.821&  1.844&  0.033&   0.30&   0.50&   0.21&    1.3&     34\\
	264& 0.5341&  4.3&    2&  2.011&  2.035&  0.070&   0.21&   0.52&   0.27&    1.1&     27\\
	460& 0.8649&  4.2&    1&  2.241&  2.268&  0.032&   0.17&   0.50&   0.32&    2.1&     80\\
	1245& 0.7875&  0.9&    7&  1.652&  1.679&  0.093&   0.59&   0.25&   0.16&    1.1&     28\\
	1494& 0.2374&  4.4&    7&  1.913&  1.930&  0.018&   0.00&   0.51&   0.49&    9.9&    380\\
	1838& 0.7450&  \nodata&    7&  1.953&  1.980&  0.065&   0.75&   0.25&   0.00&    1.0&     28\\
	1896& 0.9810&  \nodata&    7&  1.820&  1.847&  0.039&   0.14&   0.67&   0.19&    1.5&     29\\
	1970& 0.3775&  0.6&    3&  1.938&  1.960&  0.067&   0.68&   0.00&   0.32&    1.1&     22\\
	2042& 0.2362&  0.8&    7&  2.016&  2.033&  0.088&   0.00&   0.66&   0.34&    1.0&     42\\
	2087& 0.3320&  \nodata&    7&  2.060&  2.080&  0.039&   0.63&   0.16&   0.21&    3.1&    102\\
	\bf{RXJ1226.9+3332:} \\
	18& 0.7557&  3.6&    1&  2.169&  2.217&  0.029&   0.14&   0.53&   0.32&    2.3&     35\\
	91& 0.9744&  0.7&    3&  2.198&  2.223&  0.127&   0.59&   0.41&   0.00&    3.8&     16\\
	132& 0.6793&  4.7&    1&  2.174&  2.198&  0.049&   0.00&   0.75&   0.25&    7.3&     81\\
	138& 0.7136&  5.5&    1&  2.079&  2.127&  0.034&   0.18&   0.53&   0.29&    4.6&     55\\
	154& 0.6797&  8.1&    1&  2.070&  2.117&  0.072&   0.00&   1.00&   0.00&    2.3&     19\\
	185& 0.6868&  4.2&    1&  2.193&  2.241&  0.055&   0.00&   0.33&   0.67&    2.5&     35\\
	203& 0.8403&  0.6&    3&  2.383&  2.431&  0.109&   0.54&   0.30&   0.16&    2.8&     20\\
	220& 0.6882&  4.5&    1&  2.083&  2.130&  0.042&   0.00&   0.57&   0.43&    3.9&     47\\
	245& 0.9748&  4.4&    1&  2.059&  2.108&  0.045&   0.05&   0.95&   0.00&    5.8&     35\\
	247& 0.3849&  0.8&    3&  1.889&  1.908&  0.035&   0.00&   0.76&   0.24&    6.6&    172\\
	249& 0.5003&  0.6&    3&  2.349&  2.371&  0.060&   0.61&   0.00&   0.39&    5.9&     44\\
	329& 0.5883&  0.1&    3&  2.023&  2.046&  0.072&   0.63&   0.13&   0.25&    3.0&     49\\
	333& 0.7146&  1.0&    3&  1.775&  1.799&  0.056&   0.63&   0.37&   0.00&    2.9&     17\\
	347& 0.3682&  1.1&    3&  1.990&  2.009&  0.020&   0.61&   0.29&   0.10&    6.3&    103\\
	349& 0.6871&  1.9&    1&  1.964&  2.011&  0.070&   0.19&   0.49&   0.33&    5.1&     47\\
	359& 0.7026&  0.7&    3&  2.053&  2.101&  0.133&   0.61&   0.34&   0.05&    1.5&     24\\
	374& 0.3394&  1.3&    3&  2.100&  2.142&  0.093&   0.37&   0.51&   0.12&    1.4&     15\\
	386& 0.9895&  2.0&    2&  2.275&  2.324&  0.059&   0.42&   0.58&   0.00&    2.1&     15\\
	408& 0.9890&  5.3&    1&  2.095&  2.120&  0.036&   0.35&   0.50&   0.15&    5.0&     57\\
	499& 0.7326&  1.2&    3&  2.182&  2.230&  0.122&   0.00&   1.00&   0.00&    5.2&     15\\
	500& 0.4010&  1.1&    3&  2.048&  2.091&  0.055&   0.13&   0.44&   0.43&    1.2&     16\\
	523& 0.9297&  3.9&    1&  2.253&  2.278&  0.048&   0.00&   0.75&   0.25&    2.7&     30\\
	572& 0.4993&  4.8&    1&  2.082&  2.103&  0.060&   0.32&   0.56&   0.12&    2.6&     32\\
	649& 0.9254&  4.5&    1&  2.275&  2.324&  0.076&   0.21&   0.36&   0.43&    2.5&     19\\
	656& 0.9293&  4.8&    2&  2.290&  2.339&  0.043&   0.17&   0.49&   0.34&    3.8&     29\\
	739& 1.0556&  3.1&    2&  2.022&  2.071&  0.070&   0.20&   0.76&   0.04&    1.7&     14\\
		781& 0.7668&  4.8&    1&  1.933&  1.981&  0.057&   0.26&   0.74&   0.00&    6.0&     40\\
		798& 0.9636&  1.3&    3&  2.104&  2.129&  0.195&   0.59&   0.41&   0.00&    4.2&     32\\
		805& 0.6276&  1.1&    3&  2.128&  2.175&  0.066&   0.44&   0.56&   0.00&    2.4&     22\\
		824& 0.4251&  0.5&    3&  2.013&  2.057&  0.063&   0.43&   0.57&   0.00&    1.2&     21\\
		841& 0.3556&  0.9&    3&  2.083&  2.125&  0.167&   0.36&   0.64&   0.00&    1.4&     19\\
		861& 0.4208&  1.9&    2&  2.075&  2.119&  0.107&   0.16&   0.84&   0.00&    2.5&     15\\
		863& 0.7144&  2.8&    2&  1.808&  1.832&  0.036&   0.36&   0.41&   0.23&    3.9&     54\\
		872& 0.4967&  1.5&    3&  2.199&  2.221&  0.093&   0.60&   0.00&   0.40&    2.8&     11\\
		928& 0.3389&  2.0&    2&  2.020&  2.038&  0.041&   0.22&   0.69&   0.09&    7.5&     68\\
		933& 0.9645&  3.4&    1&  2.021&  2.046&  0.056&   0.15&   0.85&   0.00&    2.8&     22\\
		934& 0.4479&  1.0&    3&  2.038&  2.082&  0.071&   0.53&   0.15&   0.32&    1.4&     23\\
		960& 0.7981&  0.6&    3&  2.084&  2.133&  0.109&   0.76&   0.24&   0.00&    2.1&     22\\
		968& 0.9680&  3.5&    1&  2.302&  2.327&  0.044&   0.17&   0.55&   0.28&    4.1&     50\\
		995& 0.9611&  4.1&    1&  2.025&  2.050&  0.058&   0.69&   0.31&   0.00&    8.5&     85\\
		1001& 0.3377&  1.5&    2&  2.160&  2.202&  0.097&   0.47&   0.53&   0.00&    1.0&     10\\
		1027& 1.0588&  0.4&    3&  1.929&  1.978&  0.158&   0.39&   0.00&   0.61&    4.9&     11\\
		1080& 0.5303&  1.3&    3&  2.093&  2.115&  0.101&   0.81&   0.19&   0.00&    4.6&     47\\
		1083& 0.5341&  3.3&    2&  2.120&  2.142&  0.091&   0.55&   0.32&   0.13&    3.5&     37\\
		1103& 0.7028&  0.7&    7&  1.989&  2.012&  0.254&   0.67&   0.33&   0.00&    2.8&     22\\
		1157& 0.7669&  0.7&    3&  1.985&  2.009&  0.164&   0.80&   0.20&   0.00&    2.0&     12\\
		1175& 0.9304&  4.4&    1&  2.221&  2.246&  0.051&   0.27&   0.41&   0.32&    8.3&     95\\
		1254& 1.1236&  6.3&    2&  2.192&  2.218&  0.096&   0.20&   0.61&   0.19&    4.0&     32\\
	\enddata
	\tablenotetext{a}{Sample: (1) Passive bulge-dominated galaxies, (2) Emission line bulge-dominated
		galaxies, (3) Disk dominated galaxies, (7) Galaxies not included in the analysis. See text for 
		full description of the samples.}
	\tablenotetext{b}{Velocity dispersions corrected to a standard size aperture equivalent to a circular aperture with diameter of 3.4 arcsec at the distance of the Coma cluster (cf. \citealt{Jorgensen1995}). Velocity dispersions for galaxies in the field of MS0451.6--0305 have also been corrected for systematics, see \cite{Jorgensen2013}.}
	\tablenotetext{c}{S/N per {\AA}ngstrom in the rest frame of the galaxy. The wavelength
		interval was chosen based on the redshift of the galaxy and is typically:
		redshift $<$0.6 -- 4100-5500 {\AA};
		redshift $>$0.6 -- 3750-4100 {\AA}.
	}
\end{deluxetable*}

\begin{deluxetable*}{r rrrr rrr  rrr r r}
	\tablecaption{Line indices in the blue for the field galaxies\label{tab-speclineblue} }
	\tabletypesize{\scriptsize}
	\tablewidth{0pc}
	\tablehead{
		\colhead{Field/ID} & \colhead{H$\zeta _{\rm A}$} & \colhead{CN3883}& \colhead{CaHK}& \colhead{D4000}
		& \colhead{H$\delta _{\rm A}$}
		& \colhead{CN$_1$} & \colhead{CN$_2$} & \colhead{G4300} & \colhead{H$\gamma _{\rm A}$}
		& \colhead{Fe4383} & \colhead{C4668} 
		& \colhead{EW [\ion{O}{2}]}
	}
	\startdata
	\bf{MS0451.6--0305:} \\
	220& 3.19& 0.103& 11.90& 1.558& 3.49& \nodata& \nodata& 1.38& 0.49& 1.73& \nodata& 5.9\\
	220& 0.11& 0.005& 0.35& 0.003& 0.15& \nodata& \nodata& 0.23& 0.21& 0.29& \nodata& 0.6\\
	234& \nodata& \nodata& 7.54& \nodata& 3.48& \nodata& \nodata& 0.46& -2.00& 1.94& -1.85& \nodata\\
	234& \nodata& \nodata& 0.51& \nodata& 0.18& \nodata& \nodata& 0.17& 0.17& 0.23& 0.28& \nodata\\
	323& 2.06& 0.263& 20.92& 2.132& 0.12& 0.064& 0.094& 5.01& -4.05& 4.64& 7.68& \nodata\\
	323& 0.11& 0.005& 0.25& 0.003& 0.10& 0.002& 0.003& 0.08& 0.09& 0.11& 0.13& \nodata\\
	386& 1.59& 0.191& 20.86& 2.012& \nodata& 0.036& \nodata& 6.41& \nodata& \nodata& \nodata& \nodata\\
	386& 0.51& 0.017& 0.82& 0.012& \nodata& 0.014& \nodata& 0.36& \nodata& \nodata& \nodata& \nodata\\
	468& 3.69& 0.061& \nodata& \nodata& \nodata& \nodata& \nodata& \nodata& \nodata& \nodata& \nodata& 12.3\\
	468& 0.19& 0.008& \nodata& \nodata& \nodata& \nodata& \nodata& \nodata& \nodata& \nodata& \nodata& 1.2\\
	606& 1.67& 0.130& 20.56& 1.894& 0.53& 0.052& 0.067& 4.86& -3.51& 3.37& 6.50& \nodata\\
	606& 0.18& 0.007& 0.40& 0.004& 0.15& 0.004& 0.005& 0.12& 0.14& 0.18& 0.22& \nodata\\
	833& 1.87& 0.259& 20.31& 1.995& \nodata& \nodata& \nodata& 6.38& \nodata& 3.27& \nodata& \nodata\\
	833& 0.36& 0.015& 0.85& 0.010& \nodata& \nodata& \nodata& 0.44& \nodata& 0.66& \nodata& \nodata\\
	836& 1.54& \nodata& 7.61& \nodata& 3.70& \nodata& \nodata& 0.88& -2.07& 1.85& 2.15& \nodata\\
	836& 0.33& \nodata& 0.79& \nodata& 0.25& \nodata& \nodata& 0.27& 0.25& 0.33& 0.41& \nodata\\
	901& 5.84& 0.051& 14.06& 1.681& 8.29& \nodata& \nodata& \nodata& \nodata& \nodata& \nodata& 13.2\\
	901& 0.05& 0.002& 0.13& 0.002& 0.06& \nodata& \nodata& \nodata& \nodata& \nodata& \nodata& 0.2\\
	1584& 2.42& 0.212& 18.74& 2.072& \nodata& \nodata& \nodata& 1.12& \nodata& 11.61& \nodata& \nodata\\
	1584& 0.42& 0.017& 1.03& 0.012& \nodata& \nodata& \nodata& 0.56& \nodata& 0.55& \nodata& \nodata\\
	2032& 1.44& \nodata& 1.97& 1.098& 0.18& 0.055& \nodata& 1.32& -6.90& 0.50& 2.38& 59.5\\
	2032& 0.09& \nodata& 0.27& 0.002& 0.11& 0.003& \nodata& 0.11& 0.12& 0.15& 0.19& 0.9\\
	2230& 2.16& 0.155& 25.80& 2.118& \nodata& \nodata& \nodata& 3.34& -4.92& \nodata& \nodata& \nodata\\
	2230& 0.26& 0.011& 0.52& 0.008& \nodata& \nodata& \nodata& 0.27& 0.31& \nodata& \nodata& \nodata\\
	2491& 2.69& 0.124& 13.80& 1.840& 2.51& \nodata& \nodata& 2.87& -2.45& \nodata& \nodata& 11.9\\
	2491& 0.15& 0.007& 0.37& 0.004& 0.20& \nodata& \nodata& 0.19& 0.23& \nodata& \nodata& 1.4\\
	2561& -1.71& \nodata& 8.24& 1.648& \nodata& \nodata& \nodata& \nodata& \nodata& \nodata& \nodata& \nodata\\
	2561& 0.45& \nodata& 0.94& 0.008& \nodata& \nodata& \nodata& \nodata& \nodata& \nodata& \nodata& \nodata\\
	2657& 3.85& 0.112& 14.45& 1.498& 3.77& \nodata& \nodata& 2.85& -0.57& 3.40& 4.18& \nodata\\
	2657& 0.17& 0.008& 0.48& 0.004& 0.18& \nodata& \nodata& 0.18& 0.19& 0.27& 0.32& \nodata\\
	3005& 1.24& 0.217& 23.27& 2.205& -0.13& 0.045& 0.088& 5.18& -5.75& \nodata& \nodata& \nodata\\
	3005& 0.20& 0.007& 0.36& 0.006& 0.21& 0.005& 0.006& 0.17& 0.23& \nodata& \nodata& \nodata\\
	3521& \nodata& \nodata& \nodata& \nodata& \nodata& \nodata& \nodata& \nodata& \nodata& 2.16& 5.32& \nodata\\
	3521& \nodata& \nodata& \nodata& \nodata& \nodata& \nodata& \nodata& \nodata& \nodata& 0.22& 0.23& \nodata\\
	3610& 1.43& \nodata& 6.94& 1.185& 3.95& \nodata& \nodata& 1.43& -3.01& 1.02& -0.66& 53.1\\
	3610& 0.20& \nodata& 0.58& 0.004& 0.23& \nodata& \nodata& 0.23& 0.25& 0.39& 0.58& 7.9\\
	3625& 1.46& 0.290& 17.13& 2.232& \nodata& \nodata& \nodata& 5.43& \nodata& 3.80& \nodata& \nodata\\
	3625& 0.23& 0.009& 0.50& 0.007& \nodata& \nodata& \nodata& 0.24& \nodata& 0.30& \nodata& \nodata\\
	3635& -0.54& 0.085& 17.41& 2.009& -1.31& 0.046& 0.036& 2.29& \nodata& \nodata& \nodata& \nodata\\
	3635& 0.53& 0.019& 1.11& 0.014& 0.54& 0.015& 0.019& 0.48& \nodata& \nodata& \nodata& \nodata\\
	3697& 3.12& \nodata& 9.01& 1.455& 5.13& \nodata& \nodata& 0.95& 0.09& 2.48& 1.81& 36.0\\
	3697& 0.31& \nodata& 0.85& 0.007& 0.34& \nodata& \nodata& 0.34& 0.34& 0.52& 0.75& 5.8\\
	3792& 1.86& \nodata& 5.66& 1.551& \nodata& \nodata& \nodata& 1.85& \nodata& \nodata& \nodata& 19.7\\
	3792& 0.31& \nodata& 0.76& 0.007& \nodata& \nodata& \nodata& 0.35& \nodata& \nodata& \nodata& 0.8\\
	\bf{RXJ0152.7--1357:} \\
	103& \nodata& \nodata& \nodata& \nodata& \nodata& \nodata& \nodata& 4.15& -2.13& 2.84& \nodata& \nodata\\
	103& \nodata& \nodata& \nodata& \nodata& \nodata& \nodata& \nodata& 0.19& 0.20& 0.29& \nodata& \nodata\\
	155& 3.98& 0.085& 14.27& 1.831& 5.98& 0.028& 0.028& 2.00& 1.72& -4.24& \nodata& 5.6\\
	155& 0.28& 0.011& 0.88& 0.007& 0.25& 0.000& 0.000& 0.33& 0.36& 0.64& \nodata& 0.7\\
	193& 3.66& 0.052& 20.49& 1.775& 2.90& 0.023& 0.023& 4.59& -3.58& 5.11& 3.29& 7.5\\
	193& 0.30& 0.015& 0.68& 0.008& 0.28& 0.000& 0.000& 0.26& 0.27& 0.32& 0.34& 1.8\\
	264& 2.07& 0.127& 19.94& 1.879& -2.90& 0.093& 0.059& 3.98& -1.06& 0.82& 7.26& 11.4\\
	264& 0.50& 0.021& 0.96& 0.012& 0.47& 0.011& 0.013& 0.32& 0.30& 0.43& 0.41& 3.2\\
	460& 2.10& 0.173& 21.51& 1.923& 1.09& 0.034& 0.053& 3.35& -1.85& 2.78& 5.52& \nodata\\
	460& 0.09& 0.004& 0.22& 0.002& 0.09& 0.003& 0.004& 0.11& 0.09& 0.13& 0.21& \nodata\\
	1245& 3.81& 0.056& 8.98& 1.484& 6.06& 0.027& 0.027& 1.01& 0.36& 2.56& \nodata& 9.4\\
	1245& 0.20& 0.009& 0.54& 0.005& 0.27& 0.000& 0.000& 0.28& 0.35& 0.56& \nodata& 1.0\\
	1838& 3.66& 0.034& 6.09& 1.215& 5.12& 0.027& 0.027& 1.21& 2.14& -0.39& 5.61& \nodata\\
	1838& 0.13& 0.000& 0.45& 0.003& 0.18& 0.000& 0.000& 0.27& 0.24& 0.41& 0.51& \nodata\\
	1896& 1.51& 0.165& 20.80& 2.216& 3.09& 0.122& 0.164& 4.38& -1.89& 4.99& \nodata& \nodata\\
	1896& 0.27& 0.010& 0.76& 0.008& 0.22& 0.007& 0.008& 0.23& 0.28& 0.39& \nodata& \nodata\\
	1970& 3.51& \nodata& 13.89& \nodata& 9.87& 0.022& 0.022& 1.97& 1.09& 1.16& \nodata& \nodata\\
	1970& 0.41& \nodata& 0.98& \nodata& 0.36& 0.000& 0.000& 0.44& 0.36& 0.49& \nodata& \nodata\\
	\bf{RXJ1226.9+3332:} \\
	18& \nodata& 0.211& 21.33& 2.075& 0.28& \nodata& 0.044& 4.01& -2.25& 3.25& \nodata& 4.9\\
	18& \nodata& 0.009& 0.48& 0.006& 0.20& \nodata& 0.007& 0.21& 0.24& 0.35& \nodata& 0.8\\
	91& 2.31& \nodata& 18.94& 1.496& 1.02& \nodata& \nodata& \nodata& \nodata& \nodata& \nodata& 78.8\\
	91& 0.31& \nodata& 1.15& 0.007& 0.34& \nodata& \nodata& \nodata& \nodata& \nodata& \nodata& 18.0\\
	132& \nodata& \nodata& \nodata& \nodata& \nodata& \nodata& \nodata& 5.00& -2.59& 3.24& 7.50& \nodata\\
	132& \nodata& \nodata& \nodata& \nodata& \nodata& \nodata& \nodata& 0.09& 0.11& 0.15& 0.20& \nodata\\
	138& \nodata& 0.172& 21.05& 1.958& 1.73& \nodata& 0.029& 3.45& -2.66& 5.19& 4.90& 2.1\\
	138& \nodata& 0.006& 0.33& 0.004& 0.14& \nodata& 0.005& 0.17& 0.16& 0.21& 0.31& 0.2\\
	154& \nodata& \nodata& 24.77& \nodata& -2.32& 0.036& 0.072& 5.23& -0.62& 1.73& \nodata& \nodata\\
	154& \nodata& \nodata& 0.91& \nodata& 0.50& 0.012& 0.014& 0.46& 0.57& 0.84& \nodata& \nodata\\
	185& \nodata& 0.279& 22.33& 2.271& -0.11& 0.081& 0.105& 7.10& \nodata& \nodata& 5.63& 4.8\\
	185& \nodata& 0.013& 0.53& 0.009& 0.26& 0.007& 0.007& 0.25& \nodata& \nodata& 0.49& 0.9\\
	203& \nodata& 0.106& 9.30& 1.390& 4.63& \nodata& \nodata& \nodata& \nodata& \nodata& \nodata& 14.1\\
	203& \nodata& 0.009& 0.74& 0.005& 0.30& \nodata& \nodata& \nodata& \nodata& \nodata& \nodata& 1.5\\
	220& \nodata& 0.258& 22.91& 2.083& 1.64& \nodata& \nodata& \nodata& \nodata& 4.68& 2.94& \nodata\\
	220& \nodata& 0.009& 0.37& 0.006& 0.19& \nodata& \nodata& \nodata& \nodata& 0.24& 0.42& \nodata\\
	245& \nodata& 0.286& 20.03& 1.941& 2.07& \nodata& \nodata& \nodata& \nodata& \nodata& \nodata& \nodata\\
	245& \nodata& 0.005& 0.47& 0.004& 0.16& \nodata& \nodata& \nodata& \nodata& \nodata& \nodata& \nodata\\
	249& \nodata& \nodata& \nodata& \nodata& \nodata& \nodata& \nodata& \nodata& \nodata& \nodata& 1.26& \nodata\\
	249& \nodata& \nodata& \nodata& \nodata& \nodata& \nodata& \nodata& \nodata& \nodata& \nodata& 0.34& \nodata\\
	329& \nodata& \nodata& \nodata& \nodata& 5.05& \nodata& \nodata& \nodata& 0.23& 0.91& 1.21& \nodata\\
	329& \nodata& \nodata& \nodata& \nodata& 0.16& \nodata& \nodata& \nodata& 0.19& 0.27& 0.39& \nodata\\
	333& 5.29& 0.087& 10.68& 1.310& 5.52& \nodata& \nodata& 0.90& -2.10& \nodata& \nodata& 43.6\\
	333& 0.22& 0.012& 0.72& 0.005& 0.31& \nodata& \nodata& 0.57& 0.51& \nodata& \nodata& 1.0\\
	347& \nodata& \nodata& \nodata& \nodata& 5.01& \nodata& \nodata& 0.88& 1.30& 0.87& 1.74& \nodata\\
	347& \nodata& \nodata& \nodata& \nodata& 0.08& \nodata& \nodata& 0.09& 0.08& 0.12& 0.13& \nodata\\
	349& \nodata& 0.196& 19.20& 1.951& 2.70& \nodata& \nodata& 5.82& -2.07& 1.83& 5.28& \nodata\\
	349& \nodata& 0.008& 0.38& 0.005& 0.18& \nodata& \nodata& 0.19& 0.21& 0.26& 0.44& \nodata\\
	359& \nodata& 0.095& 8.73& 1.574& 7.14& \nodata& \nodata& -0.90& 0.15& 3.30& \nodata& 28.7\\
	359& \nodata& 0.012& 0.72& 0.006& 0.28& \nodata& \nodata& 0.46& 0.38& 0.54& \nodata& 0.9\\
	374& \nodata& \nodata& 14.13& 0.929& 3.63& \nodata& \nodata& 2.28& -0.28& 3.85& 2.26& \nodata\\
	374& \nodata& \nodata& 1.84& 0.015& 0.63& \nodata& \nodata& 0.56& 0.57& 0.82& 0.84& \nodata\\
	386& \nodata& \nodata& 7.16& 1.442& \nodata& \nodata& \nodata& \nodata& \nodata& \nodata& \nodata& 47.1\\
	386& \nodata& \nodata& 1.17& 0.007& \nodata& \nodata& \nodata& \nodata& \nodata& \nodata& \nodata& 2.3\\
	408& 3.96& 0.138& 18.42& 1.936& 4.68& \nodata& 0.048& 3.05& 0.64& 2.73& \nodata& \nodata\\
	408& 0.10& 0.005& 0.32& 0.003& 0.09& \nodata& 0.004& 0.13& 0.15& 0.23& \nodata& \nodata\\
	499& \nodata& \nodata& 12.85& 2.006& 2.53& \nodata& \nodata& \nodata& \nodata& \nodata& \nodata& \nodata\\
	499& \nodata& \nodata& 1.53& 0.015& 0.44& \nodata& \nodata& \nodata& \nodata& \nodata& \nodata& \nodata\\
	500& \nodata& \nodata& \nodata& \nodata& \nodata& \nodata& \nodata& 4.35& -0.96& 2.03& \nodata& \nodata\\
	500& \nodata& \nodata& \nodata& \nodata& \nodata& \nodata& \nodata& 0.52& 0.52& 0.72& \nodata& \nodata\\
	523& 0.84& 0.201& 19.26& 2.211& 0.30& 0.033& 0.038& 4.50& -1.89& 2.61& 5.76& \nodata\\
	523& 0.32& 0.013& 0.67& 0.009& 0.35& 0.008& 0.009& 0.30& 0.35& 0.43& 0.63& \nodata\\
	572& 4.10& 0.128& 21.23& 1.865& 3.19& \nodata& \nodata& 2.15& 2.58& 3.22& 2.09& \nodata\\
	572& 0.33& 0.015& 0.78& 0.009& 0.27& \nodata& \nodata& 0.26& 0.24& 0.34& 0.36& \nodata\\
	649& \nodata& \nodata& 21.73& 1.873& \nodata& \nodata& \nodata& 4.57& \nodata& \nodata& \nodata& \nodata\\
	649& \nodata& \nodata& 0.79& 0.008& \nodata& \nodata& \nodata& 0.40& \nodata& \nodata& \nodata& \nodata\\
	656& \nodata& 0.176& 22.99& 2.077& \nodata& \nodata& \nodata& 4.71& \nodata& \nodata& \nodata& 13.4\\
	656& \nodata& 0.011& 0.59& 0.008& \nodata& \nodata& \nodata& 0.29& \nodata& \nodata& \nodata& 3.1\\
	739& \nodata& 0.262& 21.34& 2.279& \nodata& \nodata& \nodata& \nodata& \nodata& \nodata& \nodata& 15.8\\
	739& \nodata& 0.020& 0.76& 0.015& \nodata& \nodata& \nodata& \nodata& \nodata& \nodata& \nodata& 5.6\\
	781& \nodata& \nodata& \nodata& \nodata& 3.44& \nodata& \nodata& 3.51& -0.32& 1.52& \nodata& \nodata\\
	781& \nodata& \nodata& \nodata& \nodata& 0.17& \nodata& \nodata& 0.19& 0.21& 0.37& \nodata& \nodata\\
	798& 1.07& \nodata& 2.70& 1.597& -2.21& 0.045& 0.062& 4.47& -5.21& 0.64& \nodata& 17.6\\
	798& 0.17& \nodata& 0.61& 0.004& 0.22& 0.006& 0.008& 0.23& 0.24& 0.31& \nodata& 1.5\\
	805& \nodata& \nodata& 18.57& 1.648& 3.83& \nodata& \nodata& 3.02& 1.48& \nodata& \nodata& 10.4\\
	805& \nodata& \nodata& 0.89& 0.009& 0.32& \nodata& \nodata& 0.37& 0.33& \nodata& \nodata& 0.8\\
	824& \nodata& 0.033& 16.70& 1.514& 4.39& \nodata& \nodata& 1.93& 0.70& 1.70& \nodata& 25.3\\
	824& \nodata& 0.024& 1.23& 0.011& 0.46& \nodata& \nodata& 0.39& 0.38& 0.59& \nodata& 2.4\\
	841& \nodata& \nodata& 10.87& 1.458& 3.53& \nodata& \nodata& \nodata& 1.41& 4.86& 1.52& \nodata\\
	841& \nodata& \nodata& 1.42& 0.016& 0.50& \nodata& \nodata& \nodata& 0.44& 0.61& 0.71& \nodata\\
	861& \nodata& 0.085& 11.76& 1.357& 1.76& 0.093& 0.067& 2.21& -2.51& 4.09& \nodata& 18.5\\
	861& \nodata& 0.031& 1.82& 0.014& 0.71& 0.019& 0.023& 0.59& 0.53& 0.66& \nodata& 6.0\\
	863& 2.42& 0.128& 15.89& 1.893& 3.57& \nodata& \nodata& 2.23& -1.85& 4.58& 3.65& 7.0\\
	863& 0.14& 0.007& 0.35& 0.004& 0.13& \nodata& \nodata& 0.18& 0.16& 0.22& 0.30& 1.2\\
	872& \nodata& \nodata& \nodata& \nodata& \nodata& \nodata& \nodata& 1.42& -4.98& \nodata& 2.10& \nodata\\
	872& \nodata& \nodata& \nodata& \nodata& \nodata& \nodata& \nodata& 0.73& 0.84& \nodata& 1.35& \nodata\\
	928& 1.76& 0.360& 21.04& 2.336& 3.14& \nodata& \nodata& 2.22& -0.53& 2.23& 3.52& \nodata\\
	928& 0.28& 0.017& 0.43& 0.010& 0.14& \nodata& \nodata& 0.12& 0.12& 0.18& 0.17& \nodata\\
	933& 2.49& 0.096& 20.80& 2.118& 0.70& \nodata& \nodata& 3.53& -4.28& 5.47& \nodata& \nodata\\
	933& 0.34& 0.014& 0.98& 0.010& 0.33& \nodata& \nodata& 0.42& 0.46& 0.60& \nodata& \nodata\\
	934& \nodata& \nodata& \nodata& \nodata& \nodata& \nodata& \nodata& 0.87& 1.18& 1.68& \nodata& \nodata\\
	934& \nodata& \nodata& \nodata& \nodata& \nodata& \nodata& \nodata& 0.40& 0.38& 0.53& \nodata& \nodata\\
	960& \nodata& \nodata& 9.21& 1.472& 7.12& \nodata& \nodata& \nodata& \nodata& \nodata& \nodata& 14.9\\
	960& \nodata& \nodata& 0.64& 0.006& 0.30& \nodata& \nodata& \nodata& \nodata& \nodata& \nodata& 0.8\\
	968& 2.96& 0.185& 20.90& 2.045& 1.57& 0.050& 0.068& 3.31& -1.94& 3.90& \nodata& \nodata\\
	968& 0.13& 0.006& 0.40& 0.004& 0.14& 0.004& 0.005& 0.18& 0.19& 0.28& \nodata& \nodata\\
	995& 5.79& 0.104& 16.31& 1.736& 8.61& \nodata& \nodata& 0.05& 6.82& 1.15& \nodata& \nodata\\
	995& 0.05& 0.003& 0.18& 0.002& 0.06& \nodata& \nodata& 0.12& 0.09& 0.15& \nodata& \nodata\\
	1001& \nodata& \nodata& \nodata& \nodata& 2.67& \nodata& \nodata& \nodata& 1.36& \nodata& 5.59& \nodata\\
	1001& \nodata& \nodata& \nodata& \nodata& 0.88& \nodata& \nodata& \nodata& 0.80& \nodata& 1.28& \nodata\\
	1027& \nodata& \nodata& 10.98& 1.357& \nodata& \nodata& \nodata& \nodata& \nodata& \nodata& \nodata& -8.7\\
	1027& \nodata& \nodata& 0.98& 0.010& \nodata& \nodata& \nodata& \nodata& \nodata& \nodata& \nodata& 1.9\\
	1080& 3.52& \nodata& 3.03& 1.212& 0.18& \nodata& \nodata& 0.65& -1.12& 0.85& \nodata& 68.7\\
	1080& 0.16& \nodata& 0.42& 0.003& 0.19& \nodata& \nodata& 0.19& 0.17& 0.24& \nodata& 30.4\\
	1103& 3.03& \nodata& 10.06& 1.500& 4.08& \nodata& \nodata& 1.09& -0.36& \nodata& \nodata& 20.1\\
	1103& 0.26& \nodata& 0.66& 0.006& 0.30& \nodata& \nodata& 0.49& 0.42& \nodata& \nodata& 4.0\\
	1157& 4.25& 0.116& 10.19& 1.393& 4.79& \nodata& \nodata& \nodata& 0.22& -0.55& \nodata& 20.6\\
	1157& 0.42& 0.018& 1.14& 0.009& 0.55& \nodata& \nodata& \nodata& 0.72& 1.60& \nodata& 1.8\\
	1175& 3.00& 0.190& 20.01& 1.962& 2.04& 0.028& 0.052& 3.77& 0.06& 1.21& 6.36& \nodata\\
	1175& 0.07& 0.003& 0.18& 0.002& 0.10& 0.002& 0.003& 0.09& 0.10& 0.14& 0.19& \nodata\\
	1254& \nodata& \nodata& \nodata& \nodata& 0.06& \nodata& \nodata& \nodata& \nodata& \nodata& \nodata& 5.5\\
	1254& \nodata& \nodata& \nodata& \nodata& 0.39& \nodata& \nodata& \nodata& \nodata& \nodata& \nodata& 0.2\\
	\enddata
	\tablecomments{The indices have been corrected for galaxy velocity dispersion and aperture corrected to a standard size aperture equivalent to a circular aperture with diameter of 3.4 arcsec at the distance of the Coma cluster \cite{Jorgensen1995}.
		The second line for each galaxy lists the uncertainties.}
\end{deluxetable*}

\begin{deluxetable*}{r rrrr r}
	\tablecaption{Line indices in the visible for the field galaxies\label{tab-speclinevis} }
	\tabletypesize{\scriptsize}
	\tablewidth{0pc}
	\tablehead{
		\colhead{Field/ID} & \colhead{H$\beta$} & \colhead{H$\beta _{\rm G}$} & \colhead{Mg$b$}
		& \colhead{Fe5270} & \colhead{Fe5335}
	}
	\startdata
	\bf{MS0451.6--0305:} \\
	234& -10.20& -9.78& 1.72& 0.93& 1.15\\
	234& 0.12& 0.09& 0.10& 0.10& 0.13\\
	323& 2.49& 2.58& 4.53& 3.24& 1.96\\
	323& 0.06& 0.04& 0.06& 0.09& 0.09\\
	606& 3.51& 3.25& 3.19& 2.82& 2.40\\
	606& 0.07& 0.05& 0.10& 0.12& 0.16\\
	836& -11.89& -11.87& \nodata& 0.49& 1.00\\
	836& 0.20& 0.16& \nodata& 0.21& 0.29\\
	2032& -25.51& -25.29& \nodata& -0.18& 0.29\\
	2032& 0.17& 0.14& \nodata& 0.18& 0.18\\
	2657& 0.97& 1.38& 2.10& \nodata& \nodata\\
	2657& 0.13& 0.09& 0.15& \nodata& \nodata\\
	3521& 1.11& 1.16& 2.95& 2.25& 2.47\\
	3521& 0.11& 0.07& 0.13& 0.16& 0.20\\
	3610& -16.20& -16.06& 3.22& \nodata& \nodata\\
	3610& 0.28& 0.25& 0.24& \nodata& \nodata\\
	3635& 2.13& 1.66& \nodata& \nodata& \nodata\\
	3635& 0.38& 0.26& \nodata& \nodata& \nodata\\
	3697& -7.81& \nodata& 2.35& \nodata& \nodata\\
	3697& 0.30& \nodata& 0.33& \nodata& \nodata\\
	\bf{RXJ0152.7--1357:} \\
	103& 2.32& 2.03& 2.50& 1.72& 3.10\\
	103& 0.20& 0.13& 0.21& 0.25& 0.27\\
	193& 1.03& 1.29& 3.53& 1.49& 2.43\\
	193& 0.13& 0.08& 0.15& 0.17& 0.25\\
	264& 1.83& 1.79& 4.47& 2.42& 2.97\\
	264& 0.19& 0.13& 0.28& 0.23& 0.23\\
	460& 2.28& 2.61& \nodata& \nodata& \nodata\\
	460& 0.08& 0.05& \nodata& \nodata& \nodata\\
	1245& -1.39& -2.72& \nodata& \nodata& \nodata\\
	1245& 0.31& 0.26& \nodata& \nodata& \nodata\\
	1494& \nodata& \nodata& 3.46& 2.75& 2.58\\
	1494& \nodata& \nodata& 0.02& 0.03& 0.03\\
	1838& -2.96& -3.26& -2.31& 3.74& \nodata\\
	1838& 0.33& 0.22& 0.42& 0.36& \nodata\\
	1970& -4.14& -4.07& 1.83& 1.74& 1.69\\
	1970& 0.19& 0.14& 0.18& 0.25& 0.30\\
	2042& -0.28& -0.18& \nodata& 4.34& 2.62\\
	2042& 0.25& 0.17& \nodata& 0.23& 0.25\\
	\bf{RXJ1226.9+3332:} \\
	132& 1.88& 2.04& 4.03& 3.16& 1.63\\
	132& 0.06& 0.04& 0.09& 0.12& 0.15\\
	138& 2.82& 3.06& \nodata& \nodata& \nodata\\
	138& 0.16& 0.10& \nodata& \nodata& \nodata\\
	154& 2.27& \nodata& \nodata& \nodata& \nodata\\
	154& 0.28& \nodata& \nodata& \nodata& \nodata\\
	185& 1.80& \nodata& \nodata& \nodata& \nodata\\
	185& 0.14& \nodata& \nodata& \nodata& \nodata\\
	220& 2.12& \nodata& \nodata& \nodata& \nodata\\
	220& 0.11& \nodata& \nodata& \nodata& \nodata\\
	247& \nodata& \nodata& 2.32& 2.25& 1.83\\
	247& \nodata& \nodata& 0.04& 0.06& 0.06\\
	249& -10.82& -10.50& 2.13& 2.85& 2.65\\
	249& 0.25& 0.18& 0.22& 0.26& 0.26\\
	329& -3.53& -3.17& 1.83& 2.63& 1.48\\
	329& 0.22& 0.15& 0.16& 0.27& 0.31\\
	333& -2.98& -2.24& -7.14& 1.42& 4.53\\
	333& 0.65& 0.40& 1.04& 0.58& 0.49\\
	347& -3.96& -3.81& 1.92& 0.58& 0.66\\
	347& 0.05& 0.04& 0.05& 0.06& 0.09\\
	349& 3.03& \nodata& \nodata& \nodata& \nodata\\
	349& 0.12& \nodata& \nodata& \nodata& \nodata\\
	374& -1.79& -1.30& 3.23& 0.88& 1.66\\
	374& 0.33& 0.22& 0.30& 0.30& 0.33\\
	500& 2.37& 2.14& 3.05& 3.51& 2.33\\
	500& 0.25& 0.17& 0.29& 0.31& 0.31\\
	572& 3.98& 3.63& 4.10& 1.85& 2.22\\
	572& 0.19& 0.11& 0.23& 0.28& 0.29\\
	805& \nodata& \nodata& 1.78& \nodata& \nodata\\
	805& \nodata& \nodata& 0.48& \nodata& \nodata\\
	824& -1.99& -2.51& 2.57& 2.14& 1.20\\
	824& 0.26& 0.18& 0.27& 0.28& 0.34\\
	841& -4.88& -5.04& 1.63& 1.87& 2.85\\
	841& 0.29& 0.21& 0.25& 0.26& 0.33\\
	861& -2.93& -1.43& 2.43& 1.14& 8.10\\
	861& 0.34& 0.21& 0.35& 0.31& 0.33\\
	863& 0.91& 1.00& 1.13& 1.19& 1.15\\
	863& 0.16& 0.10& 0.23& 0.17& 0.17\\
	872& -11.65& -9.75& 2.11& 7.53& 5.60\\
	872& 1.06& 0.76& 0.98& 1.13& 1.48\\
	928& 1.21& 1.51& 3.07& 2.26& 2.42\\
	928& 0.06& 0.04& 0.06& 0.05& 0.06\\
	934& -0.97& -1.13& 0.85& \nodata& \nodata\\
	934& 0.25& 0.18& 0.31& \nodata& \nodata\\
	1001& -4.03& -4.59& \nodata& 3.17& 3.82\\
	1001& 0.55& 0.42& \nodata& 0.49& 0.53\\
	1080& -11.76& -11.60& 1.36& 0.23& 1.05\\
	1080& 0.22& 0.17& 0.26& 0.24& 0.23\\
	1083& -7.07& -7.27& 2.22& -0.43& 2.03\\
	1083& 0.18& 0.14& 0.25& 0.21& 0.22\\
	1103& -2.55& -2.82& 1.33& 5.15& -0.46\\
	1103& 0.38& 0.28& 0.51& 0.45& 0.46\\
	1157& -0.21& -4.36& 2.88& \nodata& \nodata\\
	1157& 0.46& 0.43& 0.87& \nodata& \nodata\\
	\enddata
	\tablecomments{The indices have been corrected for galaxy velocity dispersion and aperture corrected to a standard size aperture equivalent to a circular aperture with diameter of 3.4 arcsec at the distance of the Coma cluster \cite{Jorgensen1995}.
		The second line for each galaxy lists the uncertainties.}
\end{deluxetable*}

\end{document}